\documentclass{ws-procs9x6}

\newcommand{\lsim}{\mathrel{\mathop{\kern 0pt \rlap
  {\raise.2ex\hbox{$<$}}}
  \lower.9ex\hbox{\kern-.190em $\sim$}}}
\newcommand{\gsim}{\mathrel{\mathop{\kern 0pt \rlap
  {\raise.2ex\hbox{$>$}}}
  \lower.9ex\hbox{\kern-.190em $\sim$}}}

\def \HM {HEIDEL\-BERG-MOSCOW~}

\def \ndbd {neutrinoless double beta decay}
\def \etal {et al.}
\def \lsim {~\mbox{${}^< \hspace*{-9pt} _\sim$}~}
\def \gsim {~\mbox{${}^> \hspace*{-9pt} _\sim$}~}

\def \znbb {$0\nu\beta\beta$}

\newcommand{\ba}[1]{\begin{eqnarray} \label{(#1)}}
\newcommand{\ea}{\end{eqnarray}}

\usepackage{times}
\usepackage{graphicx}

\begin{document}

\title{Status of Evidence for neutrinoless double
\protect\newline beta decay, and the Future: GENIUS and GENIUS-TF}

\author{H. V. Klapdor-Kleingrothaus
\footnote{\uppercase{S}pokesman of \uppercase{HEIDELBERG-MOSCOW} and 
\uppercase{GENIUS} \uppercase{C}ollaborations, 
\protect\newline \uppercase{E}-mail: klapdor@gustav.mpi-hd.mpg,
 \uppercase{H}ome-page: http://www.mpi-hd.mpg.de.non\_acc/}}

\address{Max-Planck-Institut f\"ur Kernphysik, \\ 
P.O. Box 10 39 80, D-69029 Heidelberg, Germany}

\maketitle

\abstracts{
	The first evidence for neutrinoless double beta decay has been 
	observed in the \HM experiment, which is the most sensitive 
	double beta decay experiment since ten years. 
	This is the first evidence for lepton number violation 
	and proves that the neutrino is a Majorana particle.
	It further shows that neutrino masses are degenerate. 
	In addition it puts several stringent constraints on other physics 
	beyond the Standard Model. 
	The result from the \HM experiment is consistent  
	with recent results from CMB investigations, 
	with high energy cosmic rays, with the result from the g-2 experiment 
	 and with recent theoretical work.
	It is indirectly supported by the analysis of other 
	Ge double beta experiments. 
\protect\newline The new project 
	GENIUS will cover a wide range of the parameter 
	 space of predictions of SUSY for neutralinos as cold dark matter. 
	 Further it has the potential to be a real-time detector 
	 for low-energy ($pp$ and $^7$Be) solar neutrinos. 
	 A GENIUS Test Facility has come into operation on May 5, 2003.
	This is the first time that this novel technique 
	for extreme background reduction in search for rare decays 
	is applied under the background conditions 
	of an underground laboratory.}



\section{Introduction}

	In this paper we will describe in section II 
	the recent evidence for \ndbd
	~~($0\nu\beta\beta$), found by the \HM experiment 
\cite{KK02,KK02-PN,KK-antw02,KK-BigArt02,KK02-Found}, 
	which is {\it since ten years now} the most sensitive 
	double beta experiment worldwide. 
	
	This result is
\vspace{-0.5cm}
\begin{eqnarray}
\label{5}
	{\rm T}_{1/2}^{0\nu} = (0.8 - 18.3) \times 10^{25}
	{\rm y}~~ (95\% c.l.)
\end{eqnarray}
	with best value of 
	${\rm T}_{1/2}^{0\nu} = 1.5 \times 10^{25}~ y$.
	Double beta decay is the slowest nuclear decay process 
	observed until now in nature.
	Assuming the neutrino mass mechanism to dominate the decay 
	amplitude, we can deduce
\vspace{-0.3cm}
\begin{eqnarray}
\label{6}
	\langle m_{\nu} \rangle = (0.11 - 0.56)\,eV~~ (95\% c.l.)
\end{eqnarray}
	This value we obtain using the nuclear matrix element of 
\cite{Sta90}.  
	Allowing for an uncertainty of $\pm 50\%$ of the matrix 
	elements (see 
\cite{KK02-Found,KK60Y}),
	this range widens to 
\vspace{-0.5cm}
\begin{eqnarray}
\label{7}
	\langle m_{\nu} \rangle = (0.05 - 0.84)\,eV
\end{eqnarray}

\vspace{-0.3cm}
	The result (2) and (3) determines the neutrino mass 
	scenario to be degenerate 
\cite{KK-Sark01,KK-S03-WMAP}.
	The common mass eigenvalue follows then to be 
\vspace{-0.3cm}
\begin{eqnarray}
\label{8}
	m_{com}= (0.05 - 3.2)\, eV~~ (95\%) 
\end{eqnarray}

\vspace{-0.3cm}
	If we allow for other mechanisms (see 
\cite{KK-InJModPh98,KK-SprTracts00,KK60Y,KKS-INSA02}, 
	the value given in eq. 
(\ref{6}),(\ref{7}) 
	has to be considered as an upper limit. 
	In that case very stringent limits arise for many other fields 
	of beyond standard model physics. To give an example, 
	it has been discussed recently 
\cite{Uehara02}
	that ~$0\nu\beta\beta$ decay by R-parity violating SUSY experimentally 
	may not be excluded, although this would require making R-parity 
	violating couplings generation dependent.

	We show, in section III that indirect support for the observed 
	evidence for neutrinoless double beta decay  
	evidence comes from analysis of other Ge double beta experiments 
	(though they are by far less sensitive, they yield independent 
	information on the background in the region of the expected signal).

	We discuss in sections IV and V some statistical features, 
	about which still wrong ideas are around, as well as background 
	simulations with the program GEANT4, which disprove some recent 
	criticism.

	In section VI we give a short discussion, stressing that 
	the evidence for neutrinoless double beta decay 
\cite{KK02,KK02-PN,KK-antw02,KK-BigArt02,KK02-Found}
	has been supported by various recent experimental results 
	from other fields of research 
	(see Table 
\ref{Confirm-exp}). 
	It is consistent with recent results from cosmic 
	microwave background experiments 
\cite{CMB02,WMAP03,Hannes03}.
	The precision of WMAP~~ even allows to rule out some 
	old-fashioned nuclear double beta decay matrix elements (see 
\cite{Muray03}).

\vspace{-9pt}
\begin{table}[h]
\tbl{Recent support of the neutrino mass deduced from 
	~$0\nu\beta\beta$ decay 
\protect\cite{KK02,KK02-PN,KK-antw02,KK02-Found}
	by other experiments, and by theoretical work.
\vspace*{1pt}}
{\normalsize
\newcommand{\m}{\hphantom{$-$}}
\renewcommand{\arraystretch}{.9}
\setlength\tabcolsep{8.7pt}
\begin{tabular}{|c|c|c|}
\hline
Experiment	& References	&	$m_{\nu}$ (degenerate $\nu$'s)(eV)	\\
\hline
$0\nu\beta\beta$ 	
&	\cite{KK02,KK02-PN,KK-antw02,KK02-Found}	
& 0.05 - 3.2\\
WMAP 	&	\cite{WMAP03,Hannes03}	&	$<$ 0.23, or 0.33, or 0.50\\
CMB &	\cite{CMB02}	&		$<$ 0.7 \\
Z- burst 
&	\cite{Farj00-04keV,FKR01}	& 0.08 - 1.3\\
g-2 &	\cite{MaRaid01}	&	$>$ 0.2 \\
Tritium &	\cite{Weinh-Neu00
}		& $<$2.2 - 2.8\\
$\nu$ oscillation	
& \cite{KAML02,Fogli03}			&	$>$ 0.03\\
Theory (A$_4$-symmetry)  &	\cite{BMV02}		&	$>$ 0.2 \\
Theory (identical quark &&\\
and $\nu$ mixing at GUT scale) 
&	\cite{Moh03}	
&	$>$ 0.1 \\
\hline
\end{tabular}\label{Confirm-exp}}
\vspace*{-13pt}
\end{table}


	It has been shown to be consistent with the neutrino masses 
	required for the Z-burst scenarios of high-energy cosmic rays 
\cite{FKR01,Farj00-04keV
}.
	It is consistent with a (g-2) deviating from the standard 
	model expectation 
\cite{MaRaid01}.
	It is consistent also with the limit from the tritium 
	decay experiments 
\cite{Weinh-Neu00
}
	but the allowed 95\% confidence range 
	extends down to a range which cannot be  
	covered by future tritium experiments.
	It is further strongly supported by recent theoretical work 
\cite{BMV02,Moh03}.

	Cosmological experiments like WMAP are now on the level 
	that they can seriously contribute to terrestrial research. 
	The fact that WMAP and less strictly also the tritium 
	experiments cut away the upper part of the allowed range 
	for the degenerate neutrino mass eq. 
(\ref{8}) 
	could indicate 
	that the neutrino mass eigenvalues have the same CP parity 
\cite{KK-Cp-parity03}.

	Finally we briefly comment about the possible future 
	of the field of double beta decay, and present first results 
	from GENIUS-TF which has come into operation on May 5, 2003 
	in Gran Sasso with first in world 10\,kg of naked Germanium 
	detectors in liquid nitrogen 
\cite{KK-IK-Pusk-GTF,GenTF-0012022,KK-Modul-NIM03}.



\vspace{-0.5cm}
\begin{figure}[b]
\centering{
\includegraphics*[height=5.cm,width=11cm]
{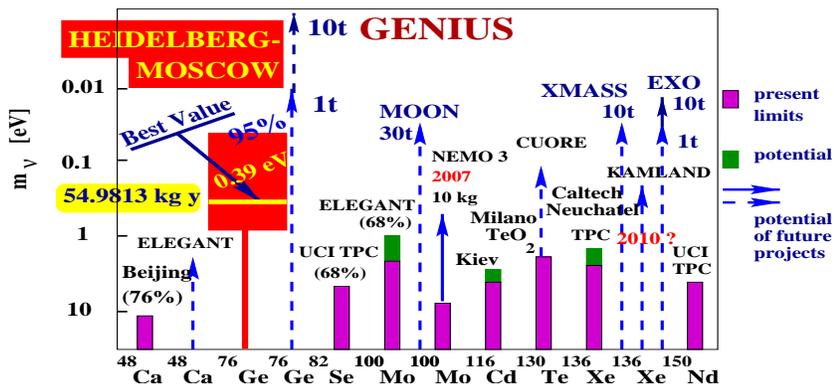}}
\caption[]{
       Present sensitivity, and expectation for the future, 
       of the most promising $\beta\beta$ experiments. 
       Given are limits for $\langle m \rangle $, except 
	for the HEIDELBERG-MOSCOW experiment where the recently 
	observed {\it value} is given (95$\%$ c.l. range and best value).
	Framed parts of the bars: present status; not framed parts: 
       future expectation for running experiments; solid and dashed lines: 
       experiments under construction or proposed, respectively. 
       For references see 
\protect\cite{KK60Y,KK02-PN,KK02-Found,KK-LowNu2,KK-NANPino00}.
\label{fig:Now4-gist-mass}}
\end{figure}


\section{Evidence for the neutrinoless decay mode}

	The status of present double beta experiments is shown in 
Fig.~\ref{fig:Now4-gist-mass}	
	and is extensively discussed in 
\cite{KK60Y}.	
	The HEIDELBERG-MOSCOW experiment using the largest source strength 
	of 11 kg of enriched $^{76}$Ge (enrichment 86$\%$) 
	in form of five HP Ge-detectors 
	is running since August 1990 
	in the Gran-Sasso underground laboratory 
\cite{KK60Y,KK02-Found,HDM01,KK02-PN,KK-StProc00,HDM97}.

\begin{figure}[b]

\vspace{-.4cm}
\centering{
\includegraphics[scale=0.3, angle=-90]
{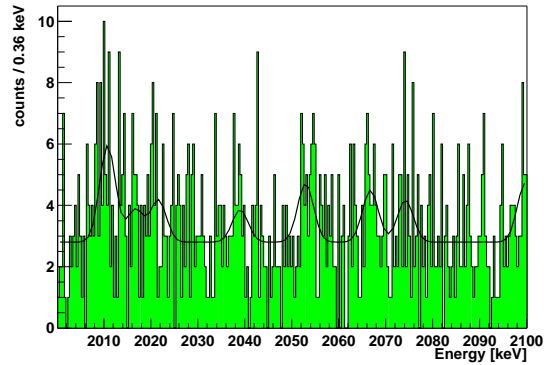}}

\vspace{-.2cm}
	\caption{The spectrum taken with the $^{76}{Ge}$ detectors 
	Nr. 1,2,3,4,5 over the period August 1990 - May 2000 
	(54.9813\,kg\,y) in the original 0.36\,keV binning, 
	in the energy range 2000 - 2100\,keV. Simultaneous 
	fit of the $^{214}{Bi}$ lines and the two high-energy 
	lines yield a probability for a line at 2039.0\,keV of 91\%.
\label{fig:100keV_sp}}
\end{figure}



	The data taken in the period August 1990 - May 2000 
	(54.9813\,kg\,y, or 723.44 mol-years) are shown in 
Fig.~\ref{fig:100keV_sp}
	in the section around the Q$_{\beta\beta}$ 
	value of 2039.006\,keV 
\cite{New-Q-2001
 }.
Fig.~\ref{fig:100keV_sp}
	is identical with 
Fig.~\ref{fig:Now4-gist-mass} 
	in 
\cite{KK02}, 
	except that we show here the original energy binning 
	of the data of 0.36\,keV. These data have been analysed 
\cite{KK02,KK02-PN,KK-BigArt02,KK02-Found}
	with various statistical methods, with the 
	Maximum Likelihood Method and 
	in particular also with the Bayesian method (see, e.g. 
\cite{KK02-Found}).

\begin{figure}[b]

\vspace{-0.5cm}
\begin{center}
\includegraphics[scale=0.2]
{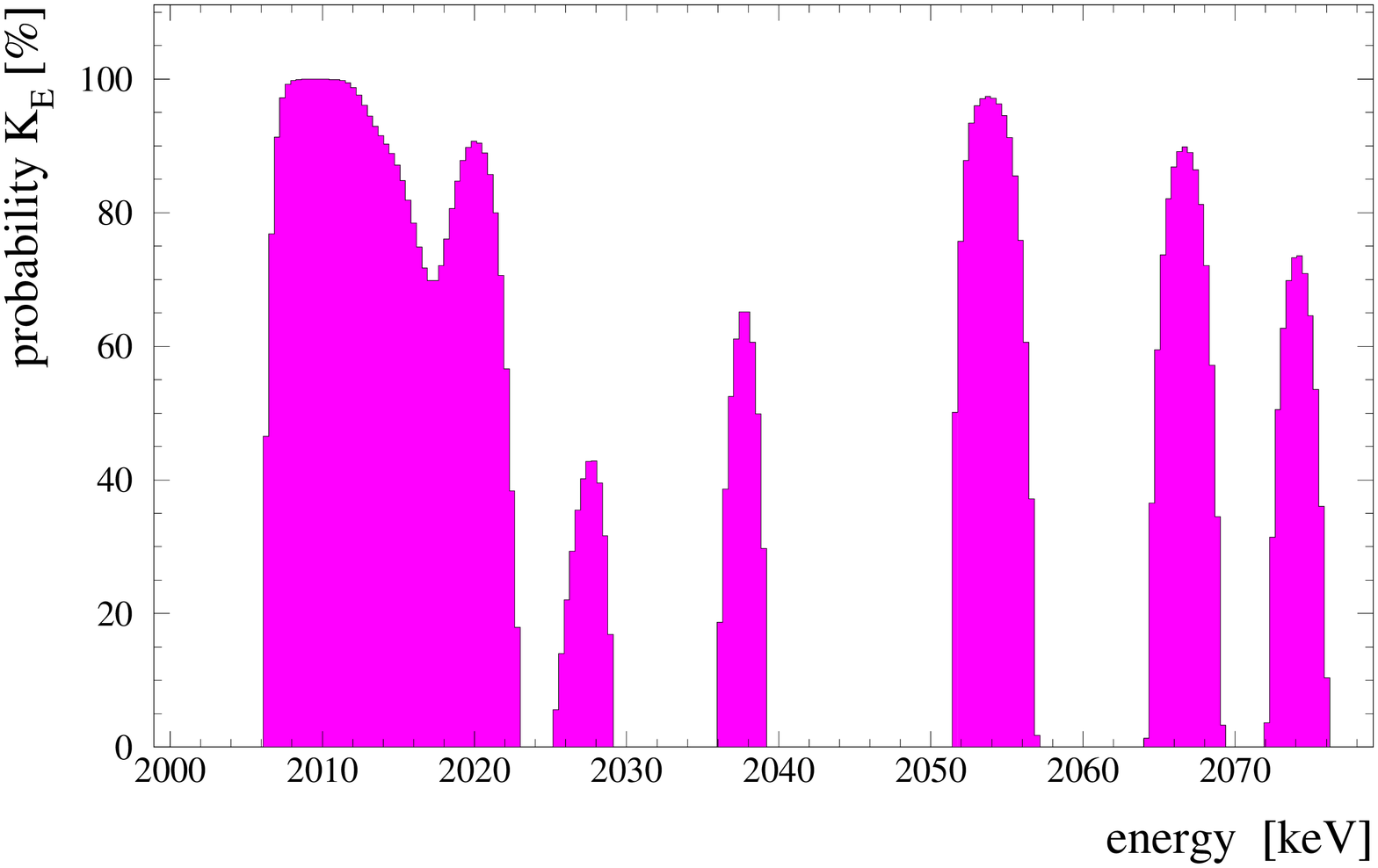}
\includegraphics[scale=0.2]
{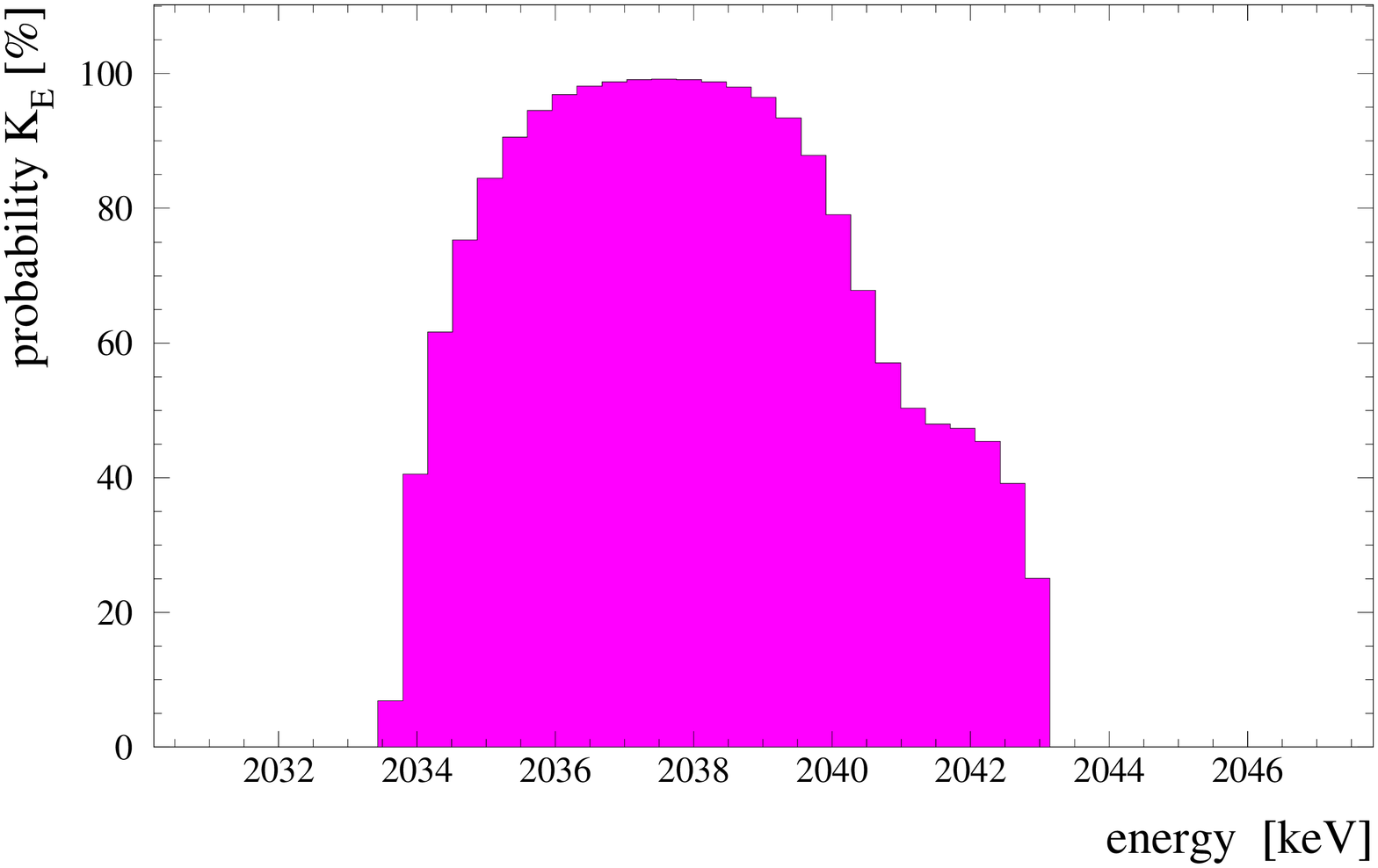}
\end{center}

\vspace{-.7cm}
	\caption{Left: 
	Probability K that a line exists at a given energy in the 
	range of 2000-2080 keV derived via Bayesian inference 
	from the spectrum shown in Fig.
\protect\ref{fig:100keV_sp}. 
	Right: 
	Result of a Bayesian scan for lines as in 
	the left part of this figure,  
	but in an energy range of $\pm 5\sigma$ around Q$_{\beta\beta}$. 
\label{fig:Bay-Chi-all-90-00-gr}}
\end{figure}


	Our peak search procedure (for details see 
\cite{KK02-PN,KK-BigArt02,KK02-Found}) 
	reproduces (see 
\cite{KK02,KK02-PN,KK-BigArt02,KK02-Found}) 
	$\gamma$-lines 
	at the positions of  known weak lines 
\cite{Tabl-Isot96} 
	from the decay of $^{214}{Bi}$ at 2010.7, 2016.7, 2021.8 
	and 2052.9 keV. 
	In addition, a line centered at 2039 keV shows up 
(see Fig. \ref{fig:Bay-Chi-all-90-00-gr}). 
	This is compatible with the Q-value 
\cite{New-Q-2001
}
	of the double beta decay process. The Bayesian analysis 
	yields, when analysing a $\pm5\sigma$ range around Q$_{\beta\beta}$ 
	(which is the usual procedure when searching for resonances 
	in high-energy physics)	a confidence level (i.e. the probability K) 
	for a line to exist at 
	2039.0 keV of 96.5 $\%$ c.l. (2.1 $\sigma$) 
(see Fig. \ref{fig:Bay-Chi-all-90-00-gr}).  
	We repeated the analysis for the same data, 
	but except detector 4, which had no muon shield 
	and a slightly worse energy resolution (46.502\,kg\,y). 
	The probability we find for a line at 2039.0 keV in this case 
	is 97.4$\%$ (2.2 $\sigma$) 
\cite{KK02,KK02-PN,KK02-Found}.
	Fitting a wide range of the spectrum yields a line 
	at 2039\,keV at 91\% c.l. (see 
Fig.\ref{fig:100keV_sp}). 

	We also applied the Feldman-Cousins method  
\cite{RPD00}.
	This method 
	(which does not use the information 
	that the line is Gaussian) finds 
	a line at 2039 keV on a confidence level of 
	3.1 $\sigma$ (99.8$\%$ c.l.).

	In addition to the line at 2039\,keV we find candidates 
	for lines at energies beyond 2060\,keV and around 2030\,keV, 
	which at present cannot be attributed. This is a task 
	of nuclear spectroscopy. 

	Important further information can be obtained from the 
	{\it time structures} 
	of the individual events. 
	Double beta events should behave as single site events 
	i.e. clearly different from a multiple scattered $\gamma$-event.  
	It is possible to differentiate between these different 
	types of events by pulse shape analysis. 
	We have developped three methods of pulse shape analysis 
\cite{HelKK00,Patent-KKHel,KKMaj99} 
	during the last seven years, one of which has been patented 
	and therefore only published recently. 	

	Installation of Pulse Shape Analysis (PSA) 
	has been performed in 1995 for the  
	four large detectors. Detector 
	Nr.5 runs since February 1995, detectors 2,3,4 since 
	November 1995 with PSA. 
	The measuring time with PSA  
	from November 1995 until May 2000 is 36.532\,kg\,years, 
	for detectors 2,3,5 it is 28.053\,kg\,y.

	In the SSE spectrum obtained 
	under the restriction that the signal simultaneously fulfills  
	the criteria of {\it all three} methods for a single site event, 
	we find again indication of a line at 2039.0\,keV (see 
\cite{KK02,KK02-PN,KK02-Found}). 

\begin{figure}[h]

\begin{center}
\includegraphics[height=4.cm,width=7cm]
{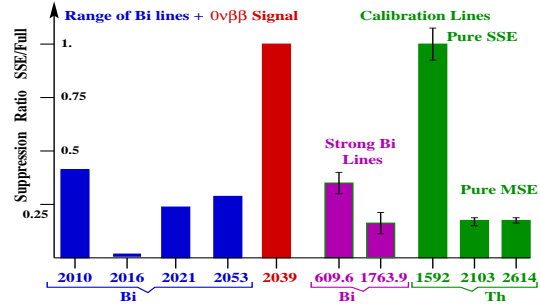} 
\end{center}

\vspace{-.3cm}
\caption{Relative suppression ratios: Remaining intensity 
	after pulse shape analysis compared to the intensity 
	in the full spectrum. Right: Result of a calibration 
	measurement with a Th source - ratio of the intensities 
	of the 1592\,keV line (double escape peak, known to be 100\% SSE), 
	set to 1. The intensities of the 2203\,keV line (single 
	escape peak, known to be 100\% MSE) are strongly reduced 
	(error bars are $\pm 1\sigma$. The same order of reduction 
	is found for the strong Bi lines occuring in our spectrum 
	- shown in this figure are the lines at 609.4 
	and 1763.9\,keV. Left: The lines in the range of weak 
	statistics around the line at 2039\,keV (shown are ratios 
	of best fit values). The Bi lines are reduced compared 
	to the line at 2039\,keV (set to 1), as to the 1592\,keV SSE Th line.}
\label{fig:Ratios}
\end{figure}

	With proper normalization concerning the running 
	times (kg\,y) of the full and the SSE spectra, 
	we see that almost the full signal remains after 
	the single site cut (best value), 
	while the $^{214}{Bi}$ lines (best values) are considerably reduced.
	We have used a $^{238}{Th}$ source to test the PSA method. 
	We find the reduction of the 2103\,keV 
	and 2614\,keV $^{228}{Th}$ lines (known to be multiple site 
	or mainly multiple site), relative to the 1592\,keV 
	$^{228}{Th}$ line (known to be single site), shown in Fig. 
\ref{fig:Ratios}. 
	This proves that the PSA method works efficiently. 
	Essentially the same reduction 
	as for the Th lines at 2103 and 2614\,keV and for the weak Bi lines 
	is found for the strong $^{214}{Bi}$ 
	lines (e.g. at 609.6 and 1763.9\,keV (Fig. 
\ref{fig:Ratios})).
	The possibility, that the single site signal is the double 
	escape line corresponding to a (much more intense!) 
	full energy peak of a $\gamma$-line, at 2039+1022=3061\,keV 
	is excluded from the high-energy part of our spectrum (see  
\cite{KK-BigArt02}).


\vspace{-0.3cm}
\section{Support of Evidence From Other Ge-Experiments 
	and From Recent Measurements With a $^{214}{Bi}$ Source}

	It has been mentioned in Section II, that by the peak 
	search procedure developped 
\cite{KK02-PN,KK02-Found}
	on basis of the Bayes and Maximum Likelihood Methods, 
	exploiting as
	important input parameters the experimental knowledge 
	on the shape and
	width of lines in the spectrum, weak lines of $^{214}$Bi 
	have been identified
	at the energies of 2010.78, 2016.7, 2021.6 and 2052.94\,keV 
\cite{KK02,KK02-PN,KK02-Found,KK-StBr02}.
	Fig. 3 shows the probability that there is a line 
	of correct width and of
	Gaussian shape at a given energy, assuming all the rest 
	of the spectrum as
	flat background (which is a highly conservative assumption).

	The intensities of these $^{214}{Bi}$ lines have been 
	shown to be consistent with other,
	strong Bi lines in the measured spectrum according 
	to the branching ratios
	given in the Table of Isotopes 
\cite{Tabl-Isot96}, 
	and to Monte Carlo simulation of the experimental setup 
\cite{KK02-Found}. 
	Note that the 2016\,keV line, as an E0 transition,
	can be seen only by coincident summing of the two successive lines
	$E=1407.98$\,keV and $E=609.316$\,keV. 
	Its observation proves that the $^{238}$U impurity from which
	it is originating, is located in the Cu cap of the detectors.
	Recent measurements of the spectrum of a $^{214}$Bi {\it source } as
	function of distance source-detector confirm this interpretation 
\cite{Oleg}.

	Premature estimates of the Bi intensities given in Aalseth et.al,
	hep-ex/\-0202018 and Feruglio et al., 
	Nucl. Phys. {B 637} (2002), 345,
	thus are incorrect, because this long-known spectroscopic
	effect of true coincident summing 
\cite{gamma} 
	has not been taken into
	account, and also no simulation of the setup has been performed (for
	details see 
\cite{KK02-Found,KK-antw02}). 

	These $^{214}$Bi lines occur also in other investigations 
	of double beta decay of Ge - and - even more important - 
	also the additional structures in
	Fig. 2, which cannot be attributed  at present, are seen in these
	other investigations.

\begin{figure}[h]

\vspace{-0.3cm}
\begin{center}
\includegraphics[width=5.cm]{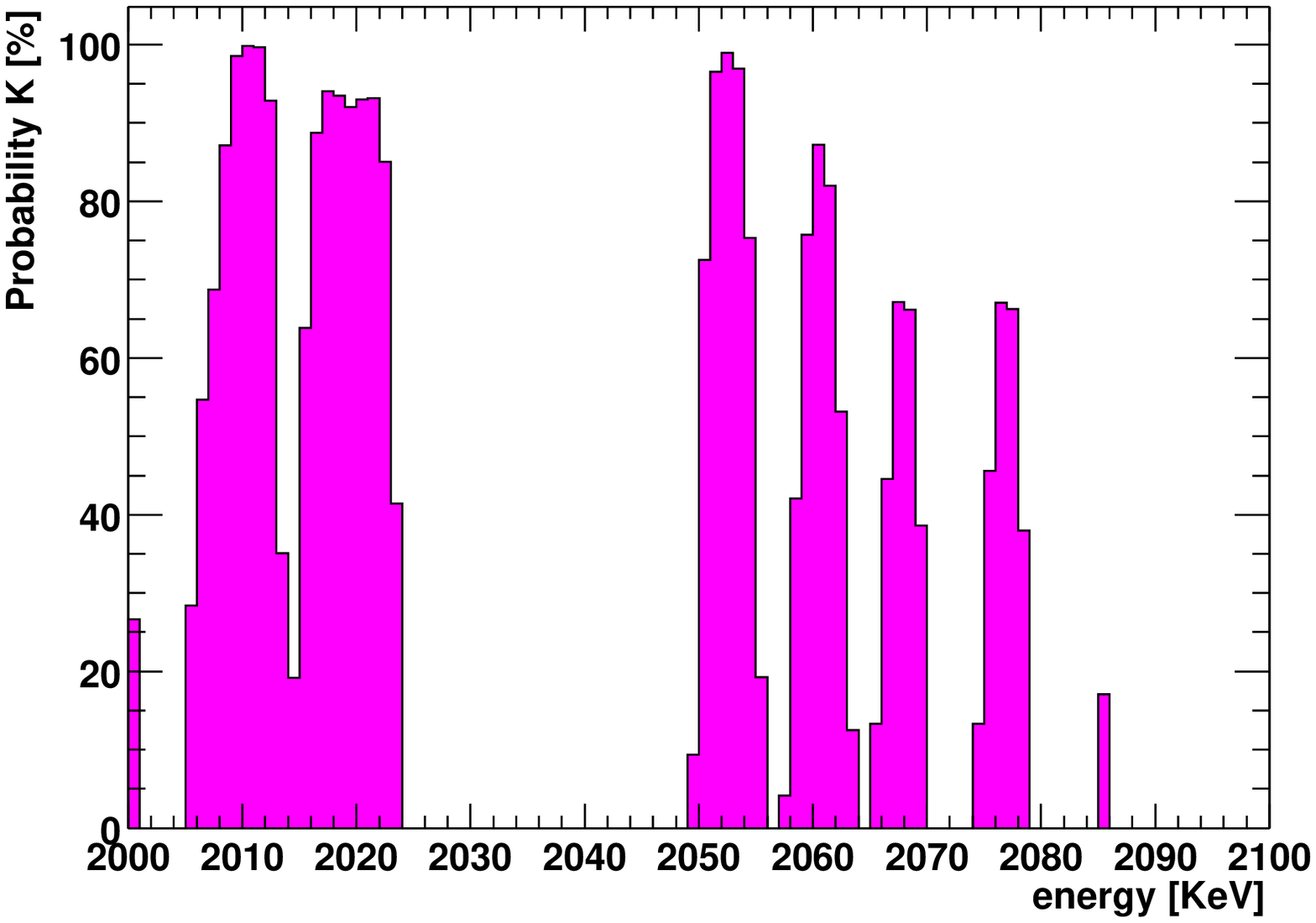}
\includegraphics[width=5.cm]{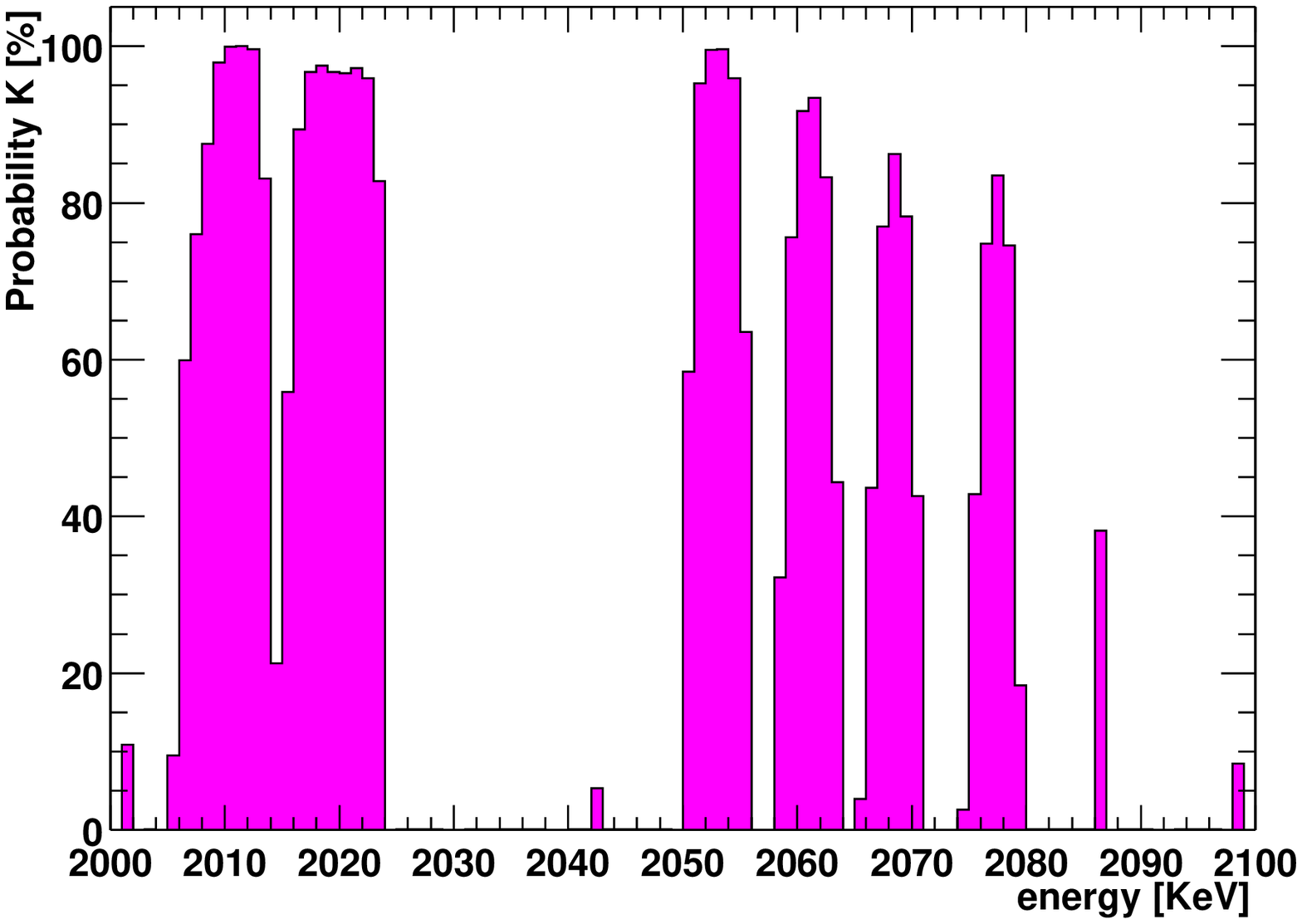}
\caption{\rm \small 
	Result of the peak-search procedure
	performed for the UCBS/LBL spectrum 
\protect\cite{Caldw91} 
	(left: Maximum Likelihood method, right: Bayes method). 
	On the y axis the probability 
	of having a line at the corresponding energy 
	in the spectrum is shown.
\label{fig:figCaldwell}}
\end{center}
\end{figure}


	There are three other Ge experiments 
	which have looked 
	for double beta decay of $^{76}$Ge. 
	First there is the experiment by Caldwell et al. 
\cite{Caldw91}, 
	using natural Germanium detectors (7.8\% abundance 
	of $^{76}$Ge, compared to 86\% in
	the HEIDELBERG-MOSCOW experiment). 
	This was the most sensitive {\it natural} Ge
	experiment. With their background a factor of 9 higher than in the
	HEIDELBERG-MOSCOW experiment and their measuring time 
	of 22.6 kg\,years,
	they had a statistics of the background by a factor of almost four
	\mbox{l a r g e r} than in the HEIDELBERG-MOSCOW experiment. 
	This gives useful
	information on the composition of the background.

	Applying the same method of peak search as used in 
Fig. 
\ref{fig:Bay-Chi-all-90-00-gr}, 
	yields indications for peaks essentially at the same energies as in
Fig. \ref{fig:Bay-Chi-all-90-00-gr} 
(see Fig. \ref{fig:figCaldwell}).  
	This shows that these peaks are not fluctuations. In particular
	it sees the 2010.78, 2016.7, 2021.6 and 2052.94\,keV $^{214}$Bi 
	lines, but a l s o  the unattributed lines at higher energies.  
	It finds, however,
	n o  line at 2039\,keV.  This is consistent with the
	expectation from the rate found in the HEIDELBERG-MOSCOW experiment.
	About 16 observed events in the latter correspond to to 0.6  expected
	events in the Caldwell experiment, because of the use of non-enriched
	material and the shorter measuring time. Fit of the Caldwell spectrum
	allowing for the $^{214}$Bi lines and a 2039\,keV line 
	yields 0.4\,events for the latter (see 
\cite{KK02-Found}). 

\begin{figure}[th]

\vspace{-0.3cm}
\begin{center}
\includegraphics[width=4.5cm]{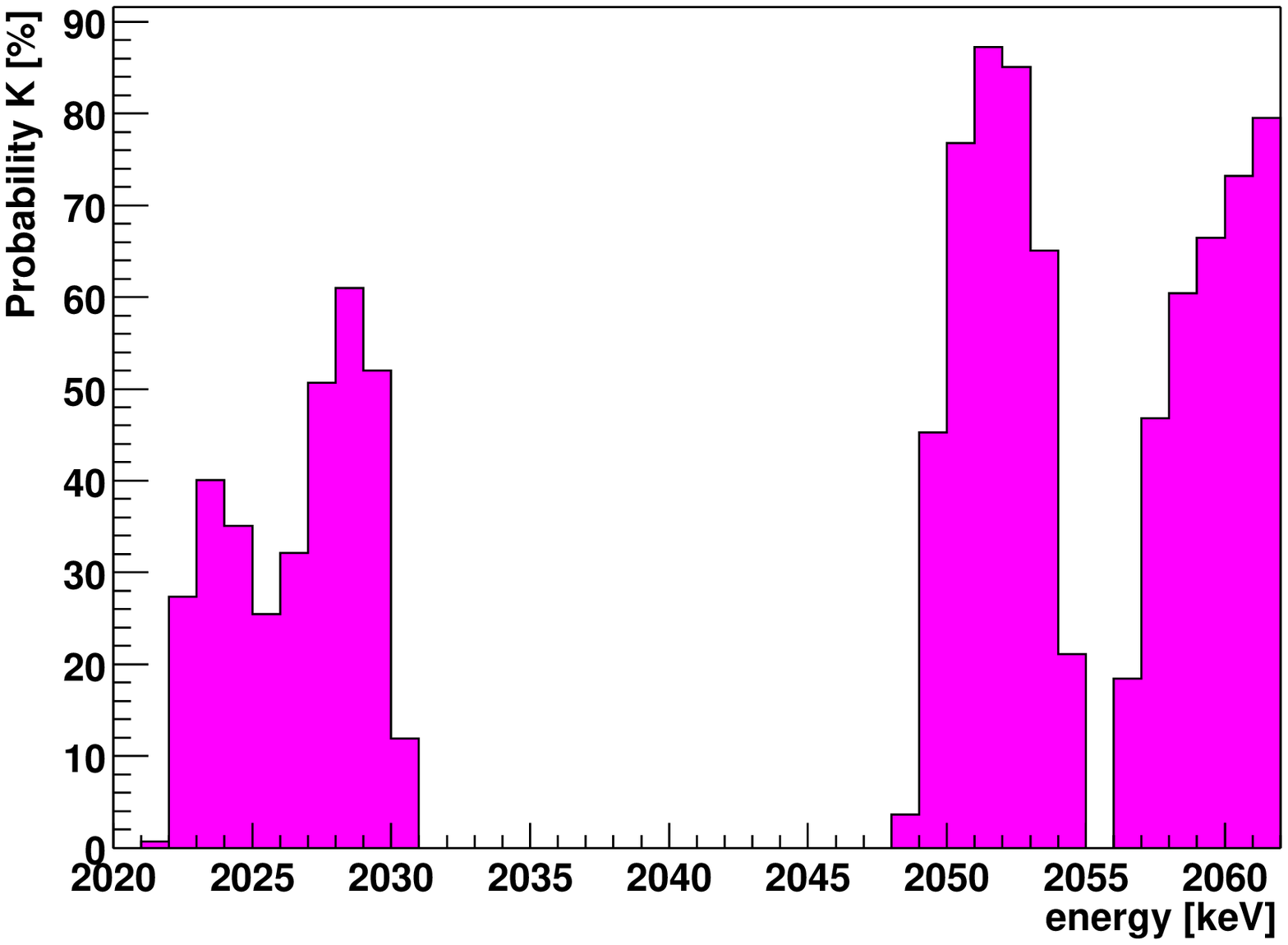}
\includegraphics[width=4.5cm]{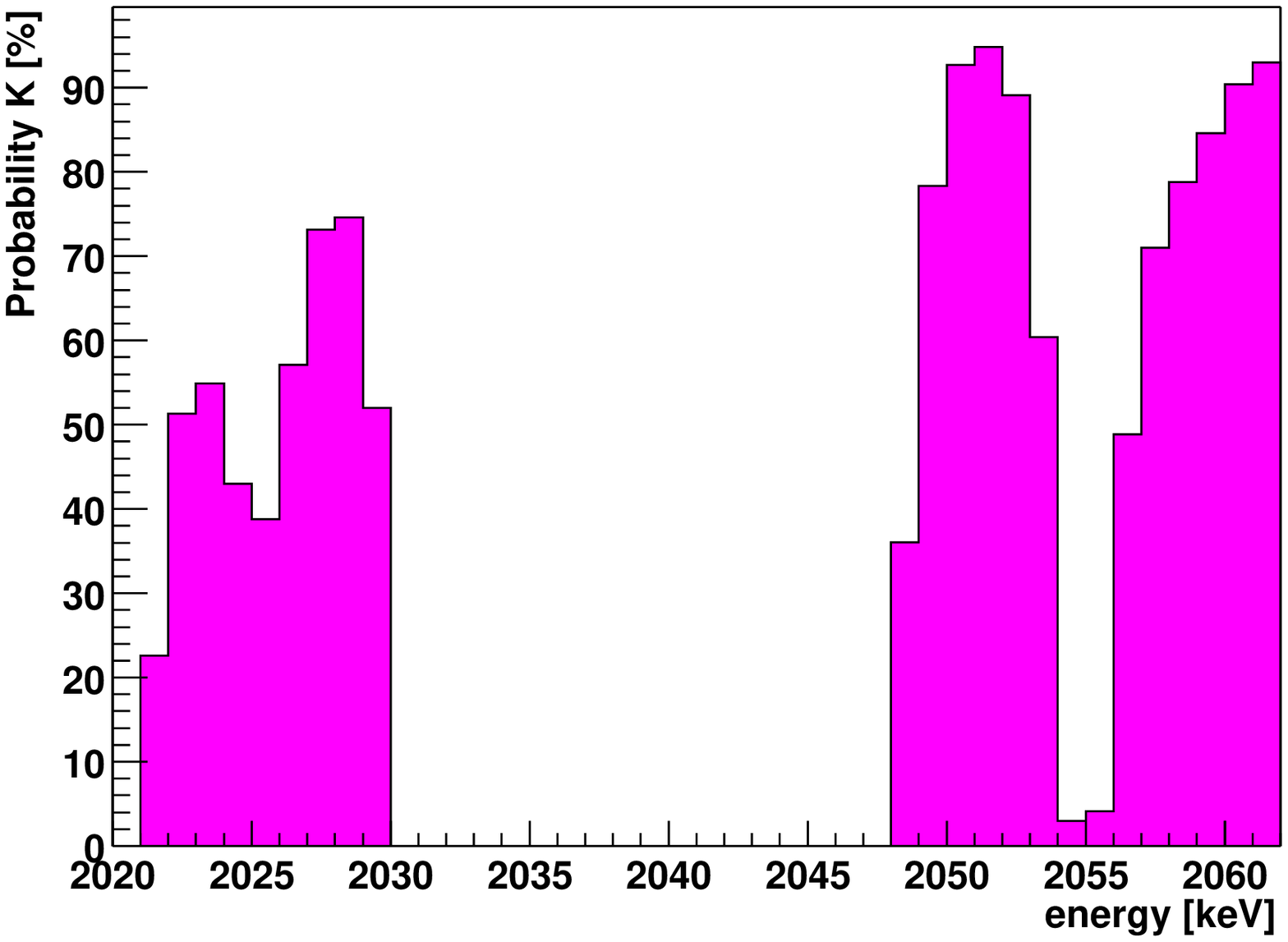}
\includegraphics[width=4.5cm]{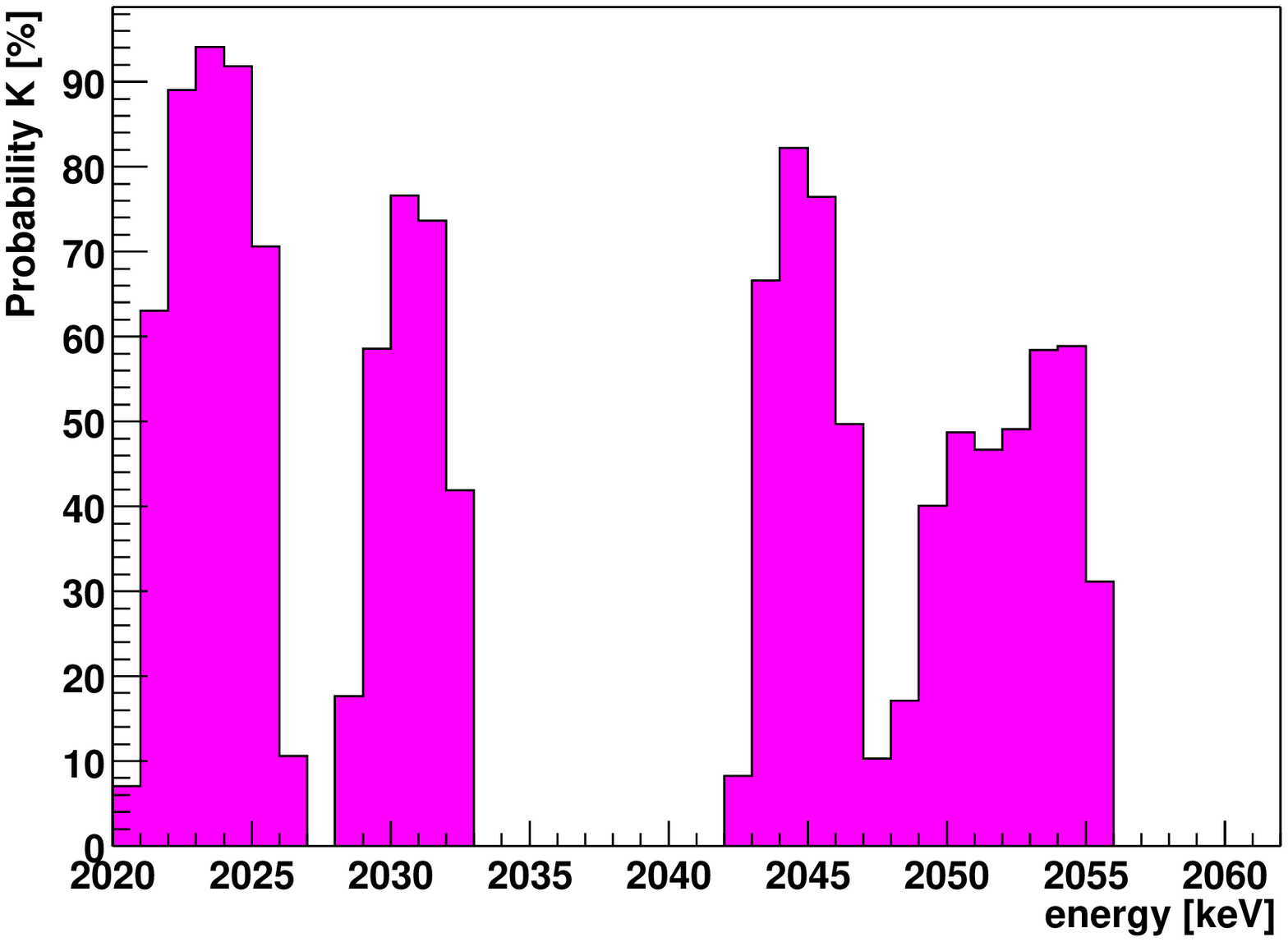}
\includegraphics[width=4.5cm]{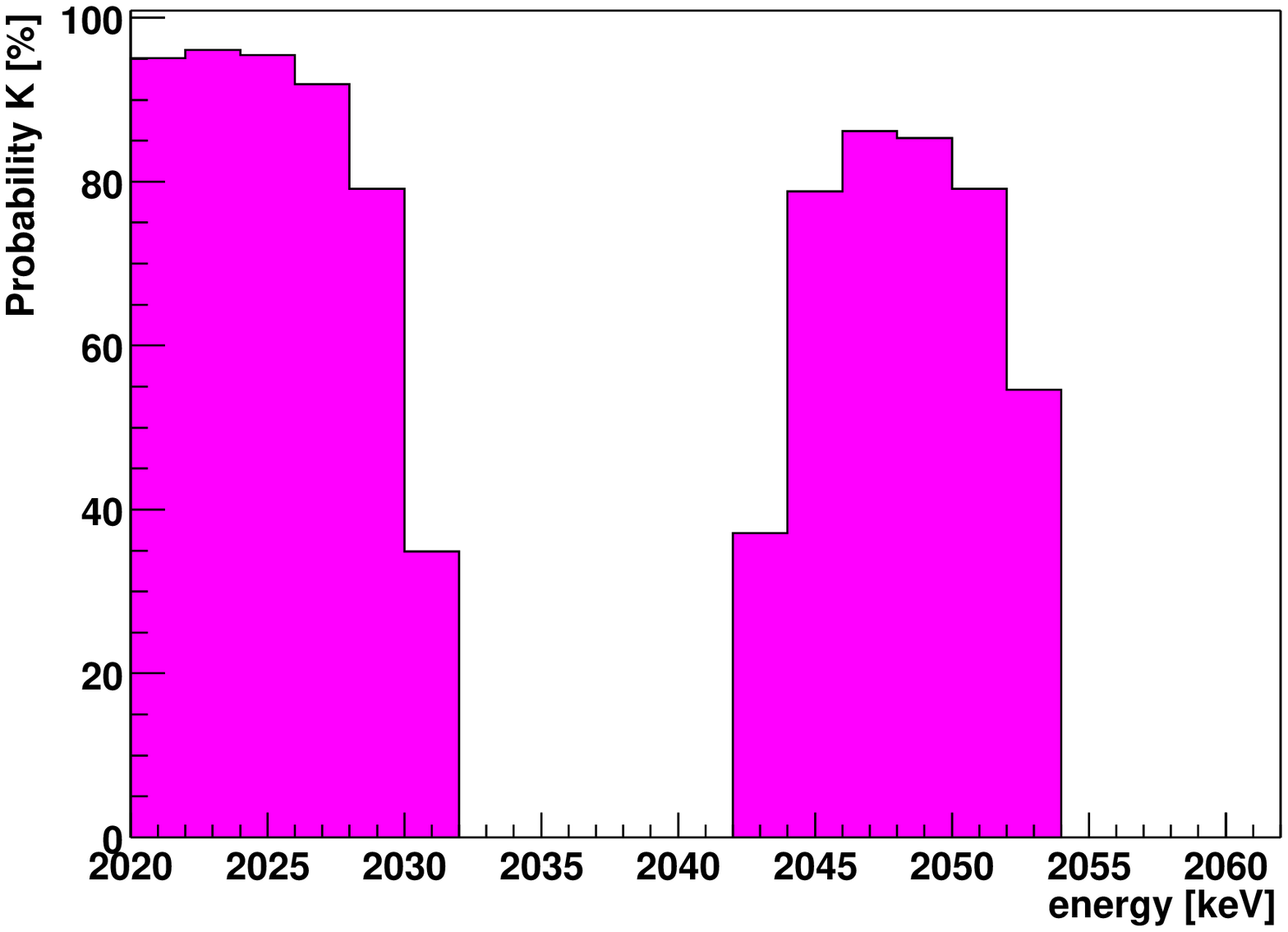}
\caption{\rm \small 
	Result of the peak-search procedure
	performed for the ITEP/YePI spectrum 
\protect\cite{Kirpichn
}
	(upper parts), and for the IGEX spectrum 
\protect\cite{DUM-RES-AVIGN-2000} 
	(lower parts).
	Left: Maximum Likelihood method, right: Bayes method. 
	On the y axis the probability 
	of having a line at the corresponding energy 
	in the specrtum is shown.
\label{fig:figITEP}}
\end{center}
\end{figure}

	The first experiment using enriched (but not high-purity) 
	Germanium 76
	detectors was that of Kirpichnikov and coworkers 
\cite{Kirpichn
}. 
	These authors show only the energy range between 2020 
	and 2064\,keV of their measured spectrum.
	The peak search procedure finds also here indications of lines
	around 2028\,keV and 2052\,keV (see Fig. 
\ref{fig:figITEP}), 
	but \mbox{n o t} any indication of a line at 2039\,keV. 
	This is consistent with the expectation, because for their low
	statistics of 2.95 kg\,y they would expect here (according to
	HEIDELBERG-MOSCOW) 0.9\,counts.

	Another experiment (IGEX) used between 6 and 8.8\,kg 
	of enriched $^{76}$Ge,
	but collected since beginning of the experiment in the early nineties
	till shutdown in 1999 only 8.8 kg\,years of statistics 
\cite{DUM-RES-AVIGN-2000}.
	The authors of 
\cite{DUM-RES-AVIGN-2000} 
	unfortunately show only the range 2020 to 2060\,keV 
	of their measured spectrum in detail. Fig. 
\ref{fig:figITEP}
	shows the result of our peak scanning 
	of this range. Clear indications are seen for
	the Bi lines  

\newpage

\vspace{-0.3cm}

\noindent
	at 2021 and 2052\,keV, but also 
	of the unidentified structure around 2030\,keV. 
	Because of the conservative assumption on the background
	treatment in the scanning procedure (see above) there 
	is no chance to see
	a signal at 2039\,keV because of the 'hole' in the background of that
	spectrum (see Fig. 1 in 
\cite{DUM-RES-AVIGN-2000}). 
	With some good will one might see, however,
	an indication of 3\,events here, consistent with the expectation of the
	HEIDELBERG-MOSCOW experiment of 2.6\,counts.


\vspace{-0.3cm}
\section{Statistical Features: Sensitivity of Peak Search, 
	Analysis Window}

\begin{figure}[h]

\vspace{-.3cm}
\begin{center}
\includegraphics[width=5.3cm]{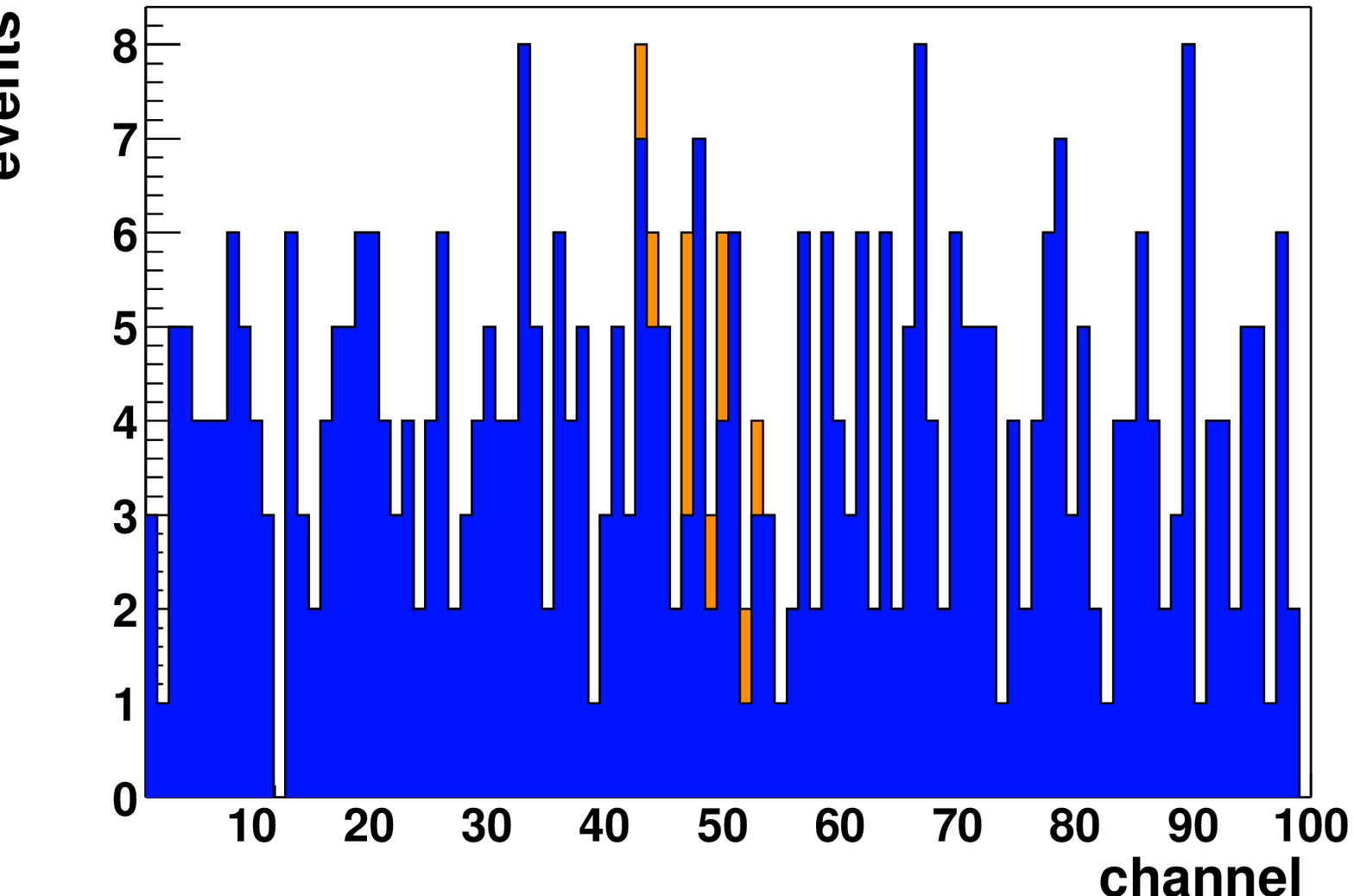}
\includegraphics[width=5.3cm]{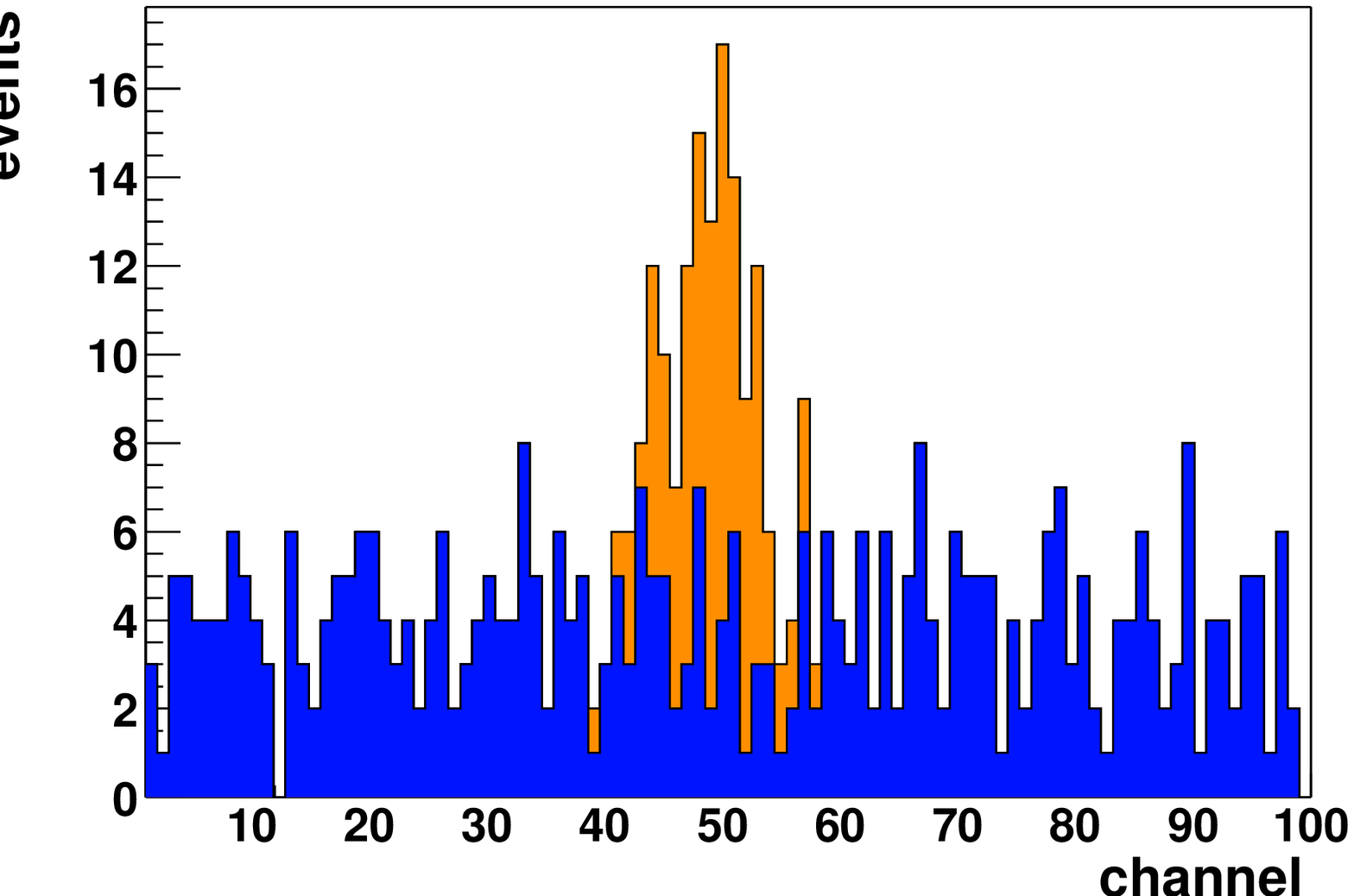}
\end{center}

\vspace{-.2cm}
\caption{\rm \small 
	Example of a random-generated spectrum with a Poisson distributed
	background with 4.0\,events per channel and a Gaussian 
	line centered in channel
	50 (line-width corresponds to a standard-deviation of $\sigma=4.0$
	channels).
	The left picture shows a spectrum with a line-intensity of 10\,events,
	the right spectrum a spectrum with a line-intensity of 100\,events.
	The background is shown dark, the events of the line bright (from 
\protect\cite{dietzdiss
,KK03}).
\label{fig:picSpecKu}}
\end{figure}

\begin{figure}[h]
\begin{center}
\includegraphics[width=5.3cm]{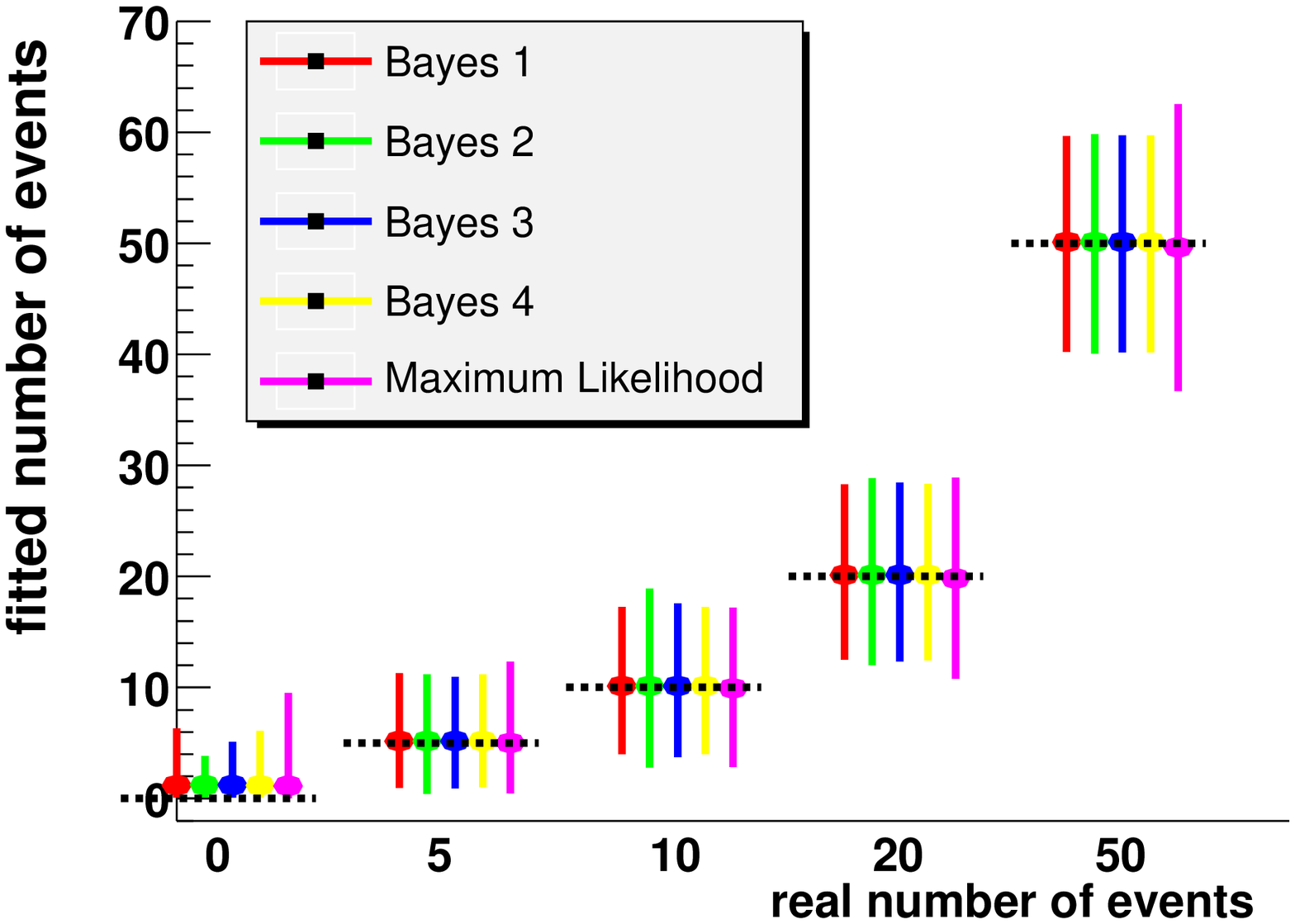}
\includegraphics[width=5.3cm]{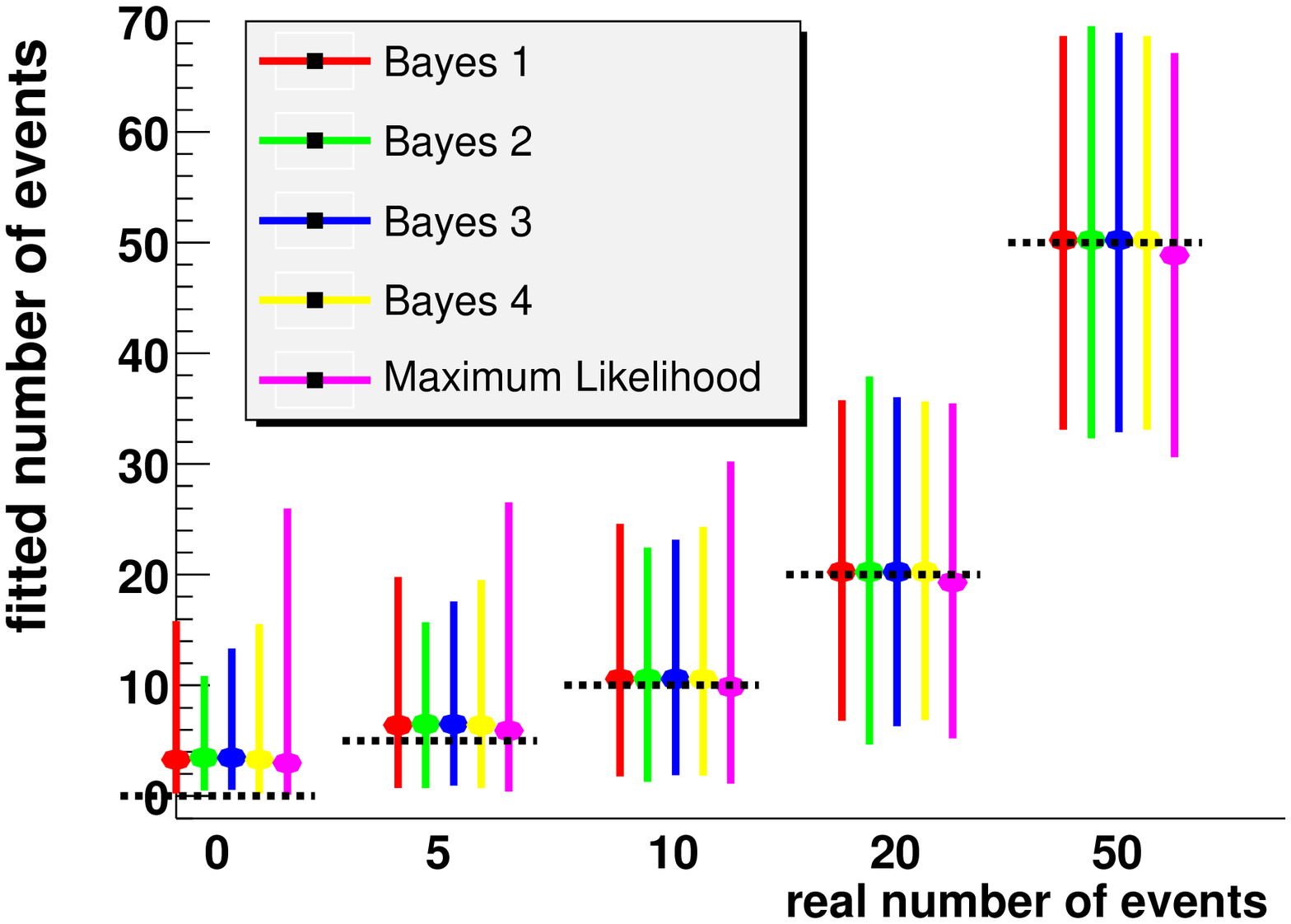}
\end{center}
\caption{\rm \small 
	Results of analysis of random-number generated spectra, 
	using Bayes and Maximum Likelihood method (the first 
	one with different prior distributions).
	For each number of events in the simulated line, 
	shown on the x-axis, 1000 random generated 
	spectra were evaluated with the five given methods.
	The analysis on the left side was performed with an Poisson
	distributed background of 0.5\,events per channel, the background for
	the spectra on the right side was 4.0\,events per channel.
	Each vertical line shows the mean value of the calculated best values
	(thick points) with the 1$\sigma$ error area.
	The mean values are in good agreement with the expected values 
	(horizontal black dashed lines) (from 
\protect\cite{dietzdiss,KK03
}).
\label{fig:picPrior}}
\end{figure}

	At this point it may be useful to demonstrate the potential 
	of the used peak search procedure. 
	Fig. 
\ref{fig:picSpecKu} 
	shows a spectrum with Poisson-generated background of
	4 events per channel and a Gaussian line with width (standard
	deviation) of 4 channels centered at channel 50, with intensity of 10
	(left) and 100 (right) events, respectively.
	Fig. 
\ref{fig:picPrior},
	shows the result of the analysis of spectra of
	different line intensity with the Bayes method (here Bayes 
	1-4 correspond to different choice of the prior distribution:
	(1) $\mu(\eta)=1$ (flat), (2) $\mu(\eta) = 1/\eta$, 
	(3) $\mu(\eta) = 1/\sqrt{\eta}$,
	(4) Jeffrey's prior) and the Maximum Likelihood Method.
	For each prior 1000 spectra have been generated with equal 
	background and equal line intensity using random number 
	generators available at CERN 
\cite{Random}.
	The average values of the best values agree (see Fig. 
\ref{fig:picPrior}) 
	very well with the known intensities also for very low 
	count rates (as in Fig.
\ref{fig:picSpecKu}, 
	left).

\begin{figure}[htb]
\begin{center}
\includegraphics[width=5.3cm]{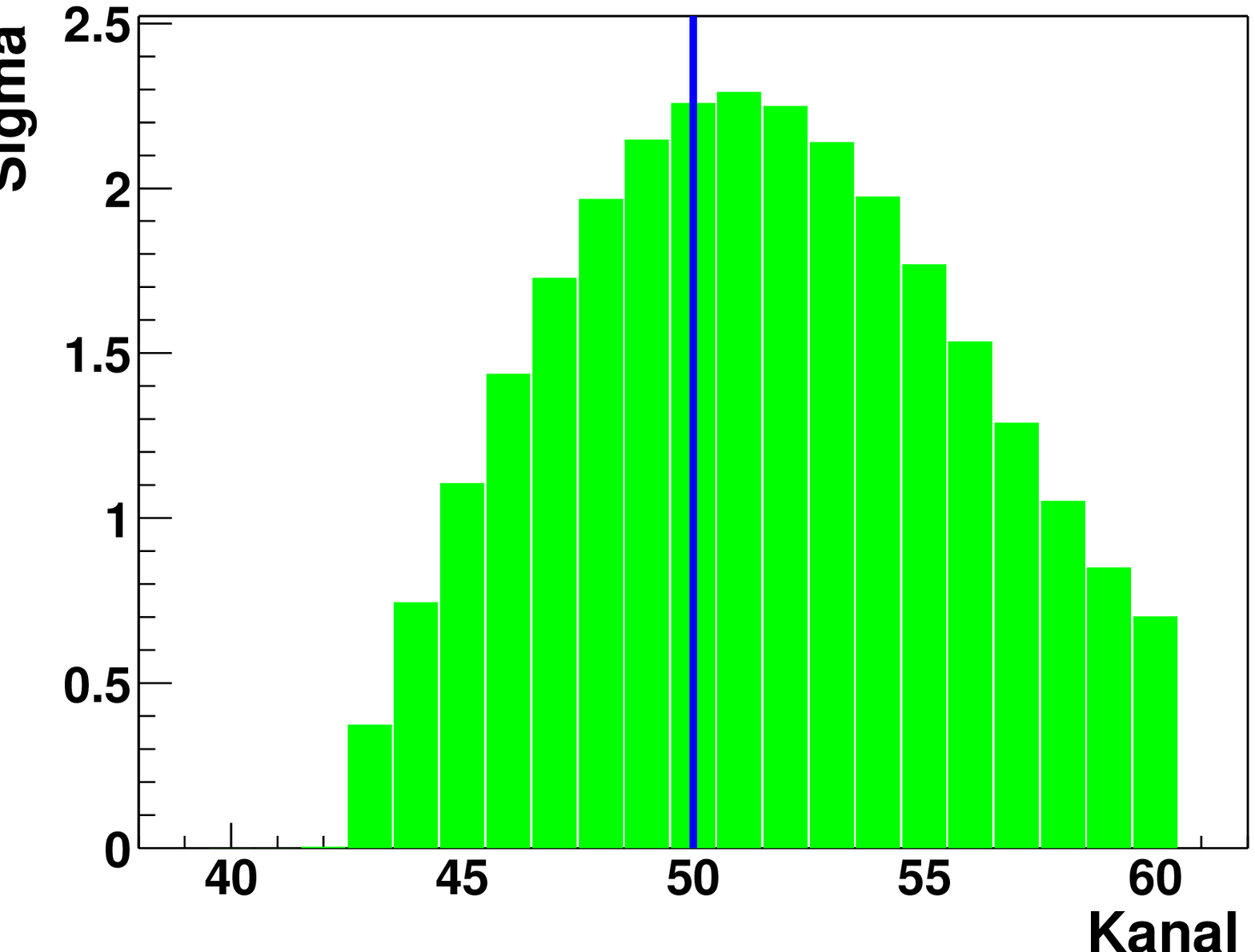}
\includegraphics[width=5.3cm]{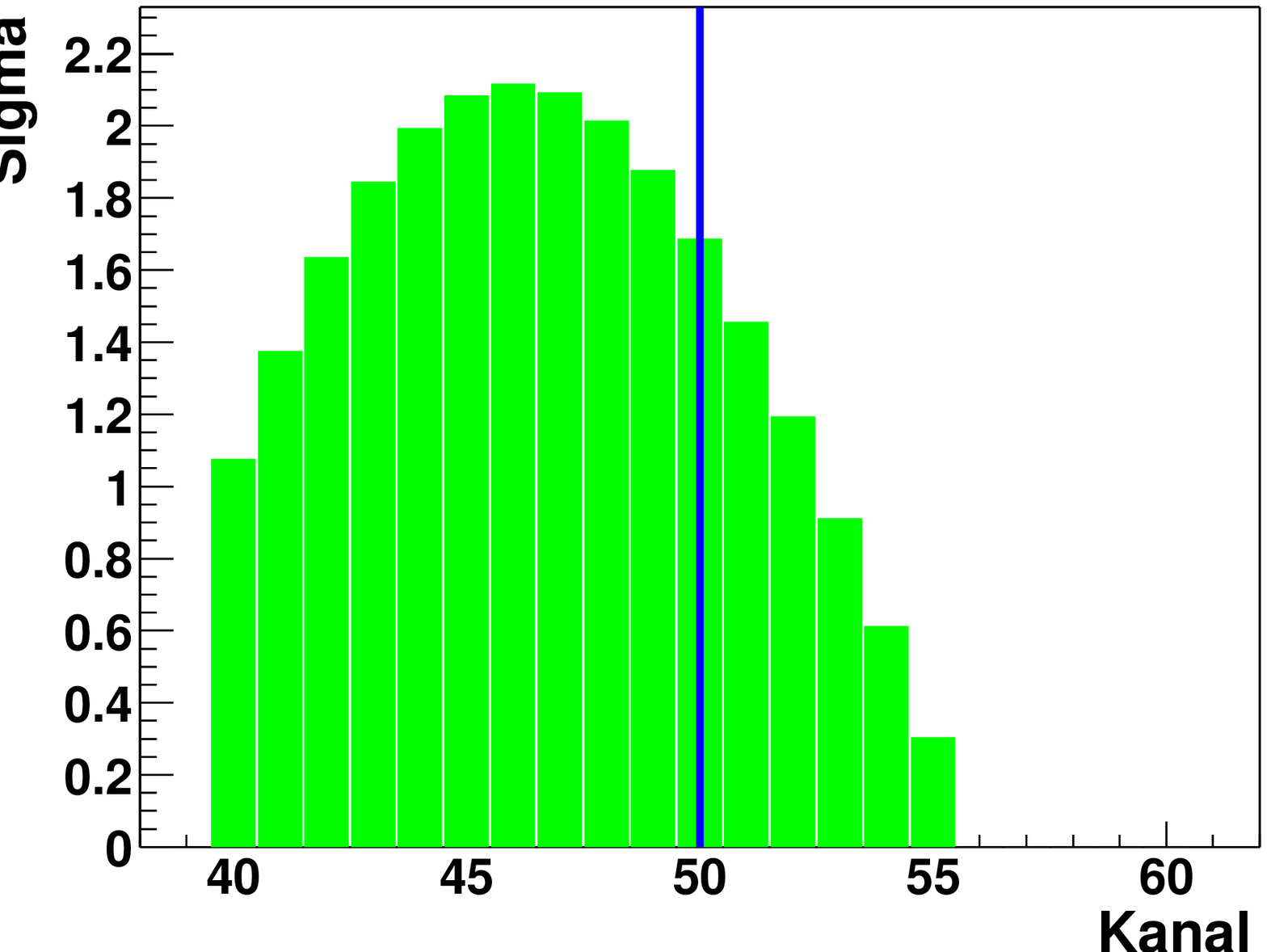}
\caption{ \rm \small 
	Two spectra with a Poisson-distributed background and a
	Gaussian line with 15 events centered in channel 50 (with a width
	(standard-deviation) of 4.0 channels) created with different 
	random numbers. 
	Shown is the result of the peak-scanning of the spectra.
	In the left picture the maximum of the probability corresponds well
	with the expected value (black line) whereas in the right
	picture a larger deviation is found. 
	When a channel corresponds  to 0.36\,keV the deviation in the right
	picture is $\sim$ 1.44\,keV (from 
\protect\cite{dietzdiss,KK03
}).
\label{fig:picWH1}}
\end{center}
\end{figure}

	In Fig. 
\ref{fig:picWH1} 
	we show two simulations of a Gaussian line of 15\,events, 
	centered at channel 50, again with width (standard deviation) 
	of 4 channels, on a Poisson-distributed background 
	with 0.5\,events/channel. The figure gives an
	indication of the possible degree of deviation of the energy of the
	peak maximum from the transition energy,  on the level of statistics
	collected in experiments like the HEIDELBERG-MOSCOW experiment 
	(here one channel corresponds to 0.36\,keV). 
	This should be considered when comparing
Figs. 
\ref{fig:Bay-Chi-all-90-00-gr},
\ref{fig:figCaldwell},\ref{fig:figITEP}.


\vspace{-.3cm}
	The influence of the choice of the energy range 
	of the analysis around $Q_{\beta\beta}$
	has been thoroughly discussed in 
\cite{KK02-PN,KK02-Found}. 
	Since erroneous ideas about this
	point are still around, 
	let us remind of the analysis given in 
\cite{KK02-PN,KK02-Found,KK03}
	which showed 
	that a reliable result is
	obtained for a range of analysis of not smaller than 35 channels
	(i.e. $\pm$18 channels) - one channel corresponding to 0.36\,keV in the
	HEIDELBERG-MOSCOW experiment.
	This is an important result, since it is of course important 
	to keep the range of analysis as  \mbox{s m a l l} 
	as possible, to avoid to include
	lines in the vicinity of the weak signal into the background.
	This unavoidably occurs when e.g. proceeding as suggested 
	in F. Feruglio et al., hep-ph/0201291 and 
	Nucl. Phys. B 637 (2002) 345-377, 
	Aalseth et. al., hep-ex/0202018 and 
	Mod. Phys. Lett. A 17 (2002) 1475,
	Yu.G. Zdesenko et. al., Phys. Lett. B 546 (2002) 206.
	The arguments given in those papers are therefore incorrect. 
	Also Kirpichnikov, who states 
\cite{Kirpichn
} 
	that his analysis finds a 2039\,keV signal in
	the HEIDELBERG-MOSCOW spectrum on a 4 sigma confidence 
	level (as we also see it, when using the Feldman-Cousins 
	method 
\cite{dietzdiss}), 
	makes this mistake 
	when analyzing the pulse-shape spectrum.


\vspace{-0.3cm}
\section{Simulation with GEANT4}

	Finally the background around $Q_{\beta\beta}$ 
	will be discussed from the side of
	simulation. A very careful new simulation of 
	the different components of
	radioactive background in the HEIDELBERG-MOSCOW experiment has been
	performed recently by a new Monte Carlo program based on GEANT4 
\cite{Doer02
,KK03}.
	This simulation uses a new event generator for simulation 
	of radioactive
	decays basing on ENSDF-data and describes the decay of arbitrary
	radioactive isotopes including alpha, beta and gamma emission 
	as well as
	conversion electrons and X-ray emission. Also included 
	in the simulation is
	the influence of neutrons in the energy range from thermal to high
	energies up to 100\,MeV on the measured spectrum. 
	Elastic and inelastic
	reactions, and capture have been taken into account, 
	and the corresponding
	production of radioactive isotopes in the setup. 
	The neutron fluxes and
	energy distributions were taken from published measurements 
	performed in
	the Gran Sasso.


\begin{figure}[t]

\vspace{-0.3cm}
\begin{center}
\includegraphics[width=12cm]{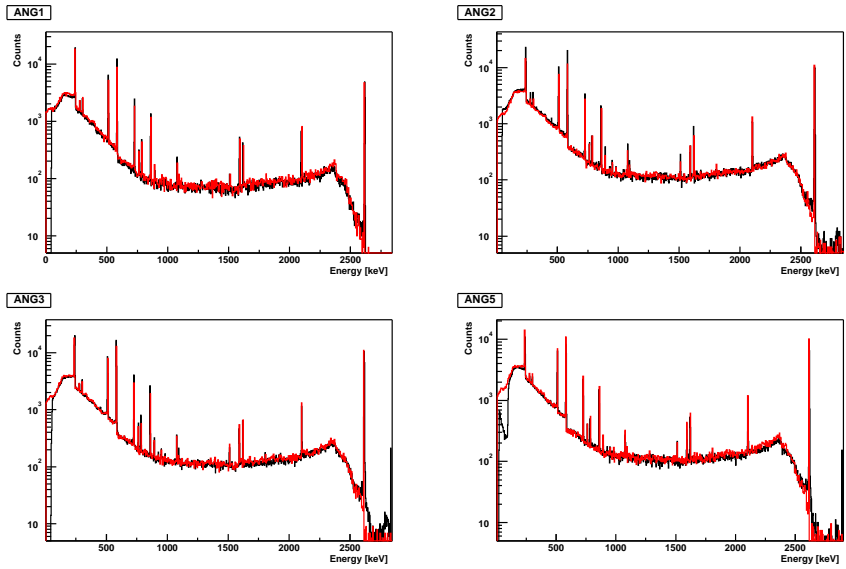}
\caption{\rm \small Comparison of the
	measured data (black line, November 1995 to April 2002) 
	and simulated
	spectrum (red line) for the detectors Nrs. 1,2,3 and 5 for a
	$^{232}$Th source spectrum.
	The agreement of simulation and measurement is excellent (from 
\protect\cite{Doer02,KK03
}).
\label{fig:picUnderTotal}}
\end{center}
\end{figure}


\begin{figure}[t]
\begin{center}
\includegraphics[width=7.cm]{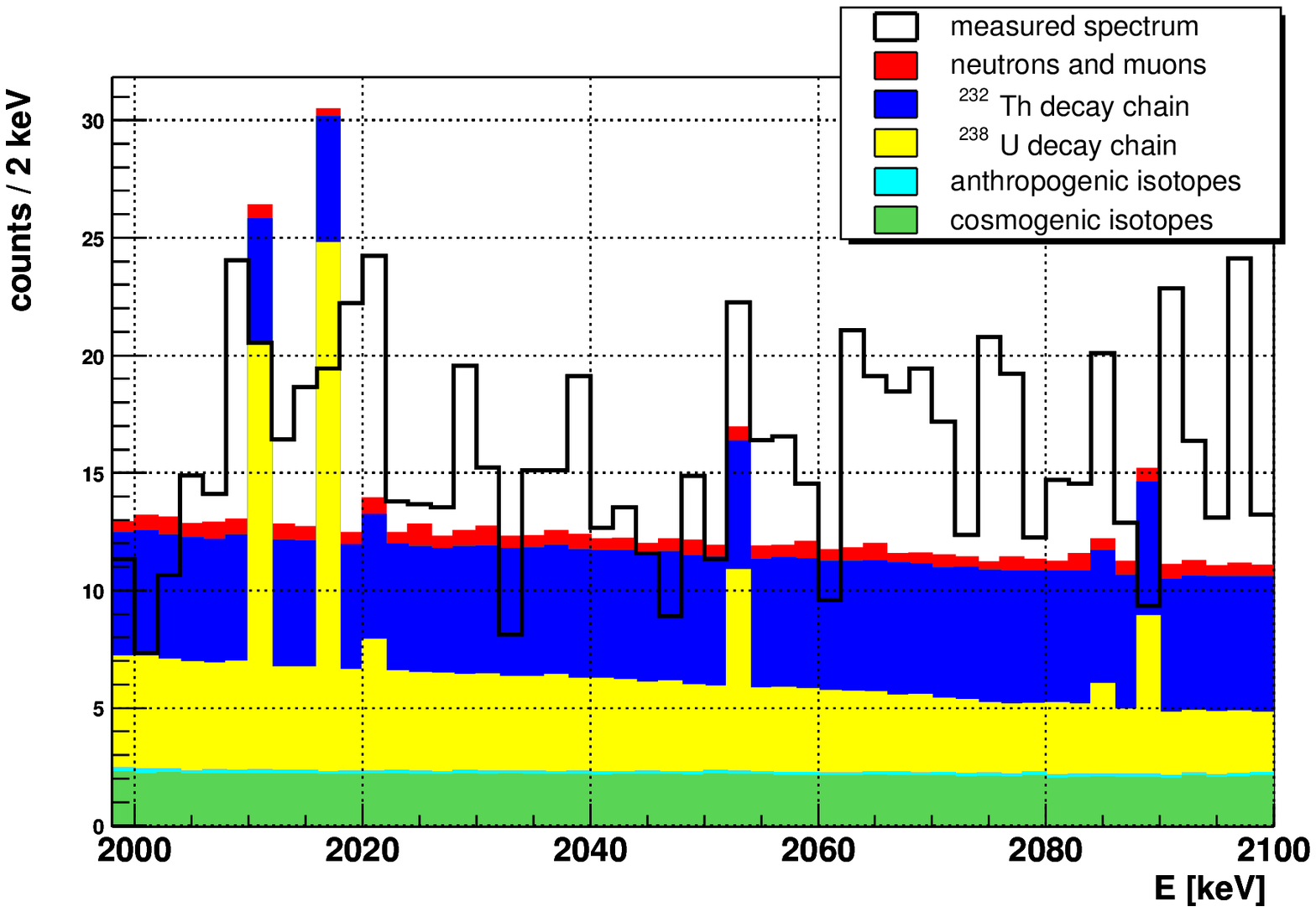}
\caption{\rm \small 
	Simulated background of the
	HEIDELBERG-MOSCOW experiment in the energy range from 2000 
	to 2100\,keV 
	with all known background components.
	The black histogram line corresponds to the measured data from
	20.11.1995 to 16.4.2002 (49.59\,kg\,y) (from 
\protect\cite{Doer02,KK03
}).
\label{fig:picUnder}}
\end{center}
\end{figure}

\noindent
	Also simulated was the cosmic 
	muon flux measured in the
	Gran Sasso, on the measured spectrum.
	To give a feeling for the quality of the simulation,
Fig. 
\ref{fig:picUnderTotal} 
	shows the simulated and the measured spectra for a $^{228}$Th 
	source spectrum for 
	as example one of our five detectors. 
	The agreement is excellent.

	The simulation of the background of the experiment 
	reproduces  a l l observed lines in the energy range 
	between threshold  (around 100\,keV) and 2020\,keV 
\cite{Doer02,KK03
}.

	Fig. 
\ref{fig:picUnder} 
	shows the simulated background in the range
	2000-2100\,keV with all  k n o w n  background components. 
	The black histogram corresponds to the measured data 
	in the period 20.11.1995 - 16.4.2002 
	(49.59\,kg\,y).

	The background around $Q_{\beta\beta}$ is according 
	to the simulations  f l a t, the
	only expected lines come from $^{214}$Bi (from the $^{238}$U 
	natural decay chain)
	at 2010.89, 2016.7, 2021.6, 2052.94, 2085.1 and 2089.7\,keV. 
	Lines from
	cosmogenically produced $^{56}$Co (at 2034.76\,keV and 2041.16\,keV), 
	half-life 77.3\,days, are not expected since the first 200\,days 
	of measurement of
	each detector are not used in the data analysis. 
	Also the potential contribution from
	decays of $^{77}$Ge, $^{66}$Ga, or $^{228}$Ac, 
	should not lead to signals visible in our
	measured spectrum near the signal at $Q_{\beta\beta}$. 
	For details we refer to  
\cite{KK03
}.



\vspace{-0.3cm}
\section{\it Proofs and disproofs}

	The result described in section 2.1 has been questioned 
	in some papers 
	(Aalseth et al, hep-ex/0202018, and in 
	Mod. Phys. Lett. A 17 92002) 1475-1478; 
	Feruglio et al., Nucl. Phys. B 637 (2002) 345; 
	Zdesenko et al., Phys. Lett. B 546 (2002) 206). 
	We think that we have shown in a convincing way that
	these claims against our results 
	are incorrect in various ways. 
	In particular the estimates of the intensities 
	of the $^{214}{Bi}$ lines in the first two papers 
	do not take into account the effect 
	of true coincidence summing, which can lead to drastic 
	underestimation of the intensities. 
	A correct estimate would also require a Monte Carlo simulation 
	of our setup, which has not been performed in the above papers.
	All of these papers, when discussing 
	the choice of the width of the search window, 
	seem to ignore the results of the statistical 
	simulations we published in 
\cite{KK02-PN,KK-antw02,KK-BigArt02,KK02-Found}.
	For details we refer to 
\cite{KK02-PN,KK-antw02,KK-BigArt02,KK02-Found,KK-StBr02}.


\vspace{-0.3cm}
\section{Discussion of results}

\subsection{\it Half-life and effective neutrino mass}

	We emphasize that we find in all analyses 
	of our spectra a line at the value of Q$_{\beta\beta}$. 
	We have shown that 
	the signal at Q$_{\beta\beta}$  
	does not originate from a background 
	$\gamma$-line. 
	On this basis we 
	translate the observed number of events 
	into half-lives for the neutrinoless double beta decay. 
	We give in Table 
\ref{Results}
	 conservatively the values obtained with 
	the Bayesian method and not those obtained with 
	the Feldman-Cousins method. 
	Also given in Table 
\ref{Results} 
	are the effective neutrino masses    
	$\langle m \rangle $ deduced using the matrix elements of 
\cite{Sta90}.


\vspace{-0.3cm}
\begin{table}[h]
\tbl{Half-life for the neutrinoless decay mode 
	and deduced effective neutrino mass 
	from the HEIDELBERG-MOSCOW experiment.
}
{\normalsize
\newcommand{\m}{\hphantom{$-$}}
\renewcommand{\arraystretch}{.9}
\setlength\tabcolsep{5.5pt}
\begin{tabular}{|c|c|c|c|c|}
\hline
$ kg\,y  $	&	Detectors		
&	${\rm T}_{1/2}^{0\nu}	{\rm~ \,y}$	
& $\langle m \rangle $ eV	&	C.L. \\
\hline
	54.9813	
&	1,2,3,4,5	
&	$(0.80 - 35.07) \times 10^{25}$	
& (0.08 - 0.54)	
& 	$95\%  ~c.l.$	\\
	&	
&	$(1.04 - 3.46) \times 10^{25}$	
& (0.26 - 0.47)	
& $68\%  ~c.l.$	\\
 		&	
&	$1.61 \times 10^{25}$	
& 0.38 	
& Best Value	\\
\hline
 	46.502	
&	1,2,3,5
&	$(0.75 - 18.33) \times 10^{25}$	
& (0.11 - 0.56)
& $95\%  ~c.l.$	\\
&	
&	$(0.98 - 3.05) \times 10^{25}$	
& (0.28 - 0.49)
& $68\% ~c.l.$	\\
&	
&	$1.50 \times 10^{25}$	
& 0.39 
& Best Value		\\
\hline
	28.053	
&	2,3,5  SSE	
&	$(0.88 - 22.38) \times 10^{25}$	
& (0.10 - 0.51)	
& $90\% ~c.l.$		\\
&	
&	$(1.07 - 3.69) \times 10^{25}$	
& (0.25 - 0.47)	
& $68\% ~c.l.$		\\
&	
&	 $1.61 \times 10^{25}$	
& 0.38
& Best Value\\	
\hline
\end{tabular}\label{Results}}

\vspace*{-13pt}
\end{table}


	We derive from the data taken with 46.502\,kg\,y 
	the half-life 
	${\rm T}_{1/2}^{0\nu} = (0.8 - 18.3) \times 10^{25}$ 
	${\rm y}$ (95$\%$ c.l.). 
	The analysis of the other data sets, shown in Table 
\ref{Results}  
	confirm this result.
	Of particular importance is that we see 
	the \znbb signal in the single site spectrum.

	The result obtained is consistent
	 with all other double beta experiments -  
	which still reach in general by far less sensitivity. 
	The most sensitive experiments following the 
	HEIDELBERG-MOSCOW experiment are the geochemical $^{128}{Te}$ 
	experiment with 
	${\rm T}_{1/2}^{0\nu} > 2(7.7)\times 10^{24}
	{\rm~ y}$  (68\% c.l.), 
\cite{manuel}
	the $^{136}{Xe}$ experiment by the DAMA group with 
	${\rm T}_{1/2}^{0\nu} > 1.2 \times 10^{24}
	{\rm~ y}$  (90\% c.l.),%
	a second enriched $^{76}{Ge}$ experiment with 
	${\rm T}_{1/2}^{0\nu} > 1.2 \times 10^{24}$ y
\cite{Kirpichn} 
	and a $^{nat}{Ge}$ experiment with 
	${\rm T}_{1/2}^{0\nu} > 1 \times 10^{24}$ y 
\cite{Caldw91}.
	Other experiments are already about a factor of 100 
	less sensitive concerning the \znbb~ 
	half-life: the Gotthard TPC experiment with $^{136}{Xe}$ yields 
\cite{Gottch} 
	${\rm T}_{1/2}^{0\nu} > 4.4 \times 10^{23}
	{\rm~ y}$  (90\% c.l.) and the Milano Mibeta cryodetector experiment 
	${\rm T}_{1/2}^{0\nu} > 1.44 \times 10^{23}
	{\rm~ y}$  (90\% c.l.).

	Another expe\-riment 
\cite{DUM-RES-AVIGN-2000}
	with enriched $^{76}{Ge}$,
	which has stopped operation in 1999 after 
	reaching a significance of 8.8\,kg\,y,
	yields (if one believes their method of 'visual inspection' 
	in their data analysis), in a conservative analysis, 
	a limit of about 
	${\rm T}_{1/2}^{0\nu} > 5 \times 10^{24}
	{\rm~ y}$  (90\% c.l.). 
	The $^{128}{Te}$ geochemical experiment 
	yields $\langle m_\nu \rangle < 1.1$ eV (68 $\%$ c.l.)
\cite{manuel},   
	the DAMA $^{136}{Xe}$ experiment 
	$\langle m_\nu \rangle < (1.1-2.9)$\,eV 
	and the $^{130}{Te}$ cryogenic experiment yields 
	$\langle m_\nu \rangle < 1.8$\,eV. 

	Concluding we obtain, with $>$ 95$\%$ probability, 
	first evidence for the neutrinoless 
	double beta decay mode. 
	As a consequence, at this confidence level, 
	lepton number is not conserved. 
	Further the neutrino is a Majorana particle. 
	If the 0$\nu\beta\beta$ amplitude is dominated by exchange 
	of a massive neutrino the effective mass 
	$\langle m \rangle $ is deduced (using the matrix elements of 
\cite{Sta90})
	to be $\langle m \rangle $ 
	= (0.11 - 0.56)\,eV (95$\%$ c.l.), 
	with best value of 0.39\,eV. 
	Allowing conservatively for an uncertainty of the nuclear 
	matrix elements of $\pm$ 50$\%$
	(for detailed discussions of the status 
	of nuclear matrix elements we refer to 
\cite{KK60Y,KK02-Found
} 
	and references therein)
	this range may widen to 
	$\langle m \rangle $ 
	= (0.05 - 0.84)\,eV (95$\%$ c.l.). 

	Assuming other mechanisms to dominate the \znbb~ decay amplitude, 
	the result allows to set stringent limits on parameters of SUSY 
	models, leptoquarks, compositeness, masses of heavy neutrinos, 
	the right-handed W boson and possible violation of Lorentz 
	invariance and equivalence principle in the neutrino sector. 
	For a discussion and for references we refer to 
\cite{KK60Y,KK-Bey97,KK-Neutr98,KK-SprTracts00,KK-NANPino00,KKS-INSA02}.

\begin{figure}[h]
\centering{
\includegraphics*[height=6.cm,width=10cm]
{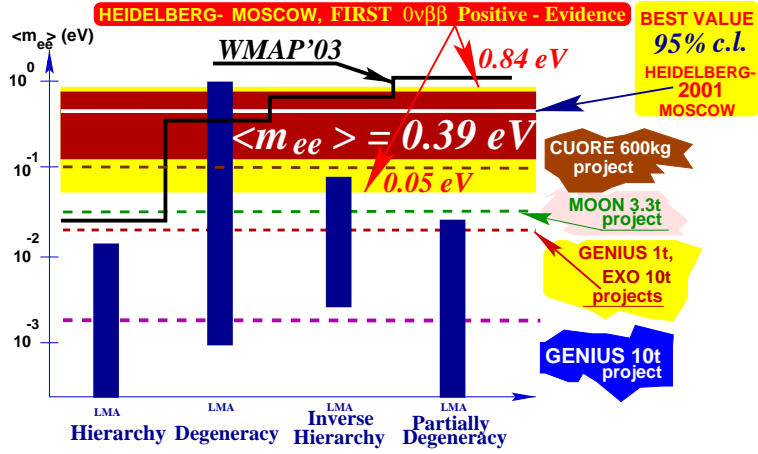}}

\vspace{.3cm}
\caption[]{
	The impact of the evidence obtained for neutrinoless 
	double beta decay (best value 
	of the effective neutrino mass 
	$\langle m \rangle$ = 0.39\,eV, 95$\%$ 
	confidence range (0.05 - 0.84)\,eV - 
	allowing already for an uncertainty of the nuclear 
	matrix element of a factor of $\pm$ 50$\%$) 
	on possible neutrino mass schemes. 
	The bars denote allowed ranges of $\langle m \rangle$ 
	in different neutrino mass scenarios, 
	still allowed by neutrino oscillation experiments (see 
\protect\cite{KK-Sark01,KK-S03-WMAP}). 
	Hierarchical models are excluded by the 
	new \znbb ~~decay result. Also shown is the exclusion line from WMAP, 
	plotted for $\sum m_{\nu} < 1.0\, eV$ 
\protect\cite{Hannes03}. 
	WMAP does not rule out any of the neutrino mass schemes. 
	Further shown are the expected sensitivities 
	for the future potential double beta experiments 
	CUORE, MOON, EXO  
	and the 1 ton and 10 ton project of GENIUS 
\protect\cite{KK60Y,KK-SprTracts00,KK-00-NOON-NOW-NANP-Bey97-GEN-prop,GEN-prop} 
	(from 
\protect\cite{KK-S03-WMAP}).
\label{fig:Jahr00-Sum-difSchemNeutr}}
\end{figure}

	With the limit deduced for the effective neutrino mass,  
	the HEIDELBERG-MOSCOW experiment excludes several 
	of the neutrino mass scenarios 
	allowed from present neutrino oscillation experiments
	(see Fig.
\ref{fig:Jahr00-Sum-difSchemNeutr}) 
	- allowing only for degenerate,  
	and  marginally still for inverse hierarchy mass scenarios 
\cite{KK-Sark01,KK-S03-WMAP}.

	The evidence for neutrinoless double beta decay has been 
	supported by various recent experimental 
	and theoretical results (see Table 1).
	Assuming the degenerate scenarios to be realized in nature 
	we fix - according to the formulae derived in 
\cite{KKPS} - 
	the common mass eigenvalue of the degenerate neutrinos 
	to m = (0.05 - 3.4)\,eV. 
	Part of the upper range is already excluded by 
	tritium experiments, which give a limit 
	of m $<$ 2.2-2.8\,eV (95$\%$ c.l.) 
\cite{Weinh-Neu00
}.
	The full range can only  partly 
	(down to $\sim$ 0.5\,eV) be checked by future  
	tritium decay experiments,  
	but could be checked by some future $\beta\beta$ 
	experiments (see next section).
	The deduced best value for the mass 
	is consistent with expectations from experimental 
	$\mu ~\to~ e\gamma$
	branching limits in models assuming the generating 
	mechanism for the 
	neutrino mass to be also responsible 
	for the recent indication for as anomalous magnetic moment 
	of the muon
\cite{MaRaid01}.  
	It lies in a range of interest also for Z-burst models recently 
	discussed as explanation for super-high energy cosmic ray events 
	beyond the GKZ-cutoff 
\cite{Farj00-04keV,
FKR01} 
	and requiring neutrino masses in the range (0.08 - 1.3)\, eV. 
	A recent model with underlying A$_4$ symmetry for 
	the neutrino mixing matrix also leads to degenerate 
	neutrino masses $>$ 0.2\, eV, consistent with the present result 
	from \znbb~ decay 
\cite{Ma-DARK02,BMV02}. 
	The result is further consistent with the theoretical paper of 
\cite{Moh03}.
	Starting with the hypothesis that quark and lepton mixing are
	identical at or near the GUT scale, Mohapatra \etal 
~\cite{Moh03} show
	that the large solar and atmospheric neutrino mixing angles can be
	understood purely as result of renormalization group evolution, if
	neutrino masses are quasi-degenerate (with same CP parity). 
	The common 
	Majorana neutrino mass then must be, in this model, 
	larger than 0.1 eV.

	For WMAP a limit on the total neutrino masses of 
\vspace{-0.2cm}
\begin{equation}
m_s = \sum m_i < 0.69 ~{\rm eV~~~~~~~ at~ } 95\%~~ {\rm c.l.},
\end{equation}
	is given by the analysis of ref. 
\cite{WMAP03}. 
	It has been shown, however, that this limit may 
	not be very realistic. 
	Another analysis shows that this limit on the total 
	mass should be 
\cite{Hannes03}
\vspace{-0.3cm}
\begin{equation}
m_s = \sum m_i < 1.0 ~{\rm eV~~~~~~~ at }~ 95\% ~~{\rm c.l.} 
\end{equation}
	The latter analysis also shows, that four 
	generations of neutrinos are still allowed and in the
	case of four generations the limit on the total mass is 
	increased to $1.38$ eV. If there is
	a fourth neutrino with very small mass, then the limit 
	on the total mass of the three neutrinos is even further weakened 
	and there is essentially no
	constraint on the neutrino masses. 
	In our Fig. 
\ref{fig:Jahr00-Sum-difSchemNeutr}
	we show the contour line for WMAP assuming 
	$\sum m_i < 1.0 ~{\rm eV}$.
 	
	Comparison of the WMAP results with the effective mass 
	from double beta decay rules out completely (see 
\cite{Muray03})
	a 15\,years old old-fashioned 
	nuclear matrix element of double beta decay, used
	in a recent analysis of WMAP  
\cite{Vogel1}. 
	In that calculation of the nuclear matrix element there was 
	not included a realistic nucleon-nucleon
	interaction, which has been included by all other calculations of the
	nuclear matrix elements over the last 15\,years. 

	As mentioned in section 1 the results from double beta decay 
	and WMAP {\it together} may indicate 
\cite{KK-Cp-parity03} 
	that the neutrino mass eigenvalues have indeed the same CP parity, 
	as required by the model of 
\cite{Moh03}. 

	The range of $\langle m \rangle $ fixed in this 
	work is, already now, in the range to be explored 
	by the satellite experiments MAP and PLANCK 
\cite{KKPS,WMAP03,Hannes03}. 
	The limitations of the information from WMAP are seen in Fig. 
\ref{fig:Jahr00-Sum-difSchemNeutr}, 
	thus results of PLANCK are eagerly awaited.

	The neutrino mass deduced leads to 0.002$\ge \Omega_\nu h^2 \le$
	0.1 and thus may allow neutrinos to still play 
	an important role as hot dark matter in the Universe 
\cite{KK-LP01}.


\vspace{-0.3cm}
\section{Future of $\beta\beta$ experiments - GENIUS and other proposals}

	With the HEIDELBERG-MOSCOW experiment, the era of the small smart 
	experiments is over. 
	New approaches and considerably enlarged experiments 
	(as discussed, e.g. in 
\cite{KK-InJModPh98,KK-Bey97,KK60Y,KK-Neutr98,KK-00-NOON-NOW-NANP-Bey97-GEN-prop,GEN-prop,KK-NOW00,KK-LP01})
	will be required in future 
	to fix the neutrino mass with higher accuracy. 
	
	Since it was realized in the HEIDELBERG-MOSCOW experiment, 
	that the remaining small background is coming from the material 
	close to the detector (holder, copper cap, ...), 
	elimination of {\it any} material close to the detector 
	will be decisive. Experiments which do not take this 
	into account, like, e.g. CUORE 
	and MAJORANA 
	will allow at best only rather limited steps in sensitivity. 
	Furthermore there is the problem in cryodetectors that they 
	cannot differentiate between a $\beta$ and a $\gamma$ signal, 
	as this is possible in Ge experiments.

	Another crucial point is the energy resolution, 
	which can be optimized {\it only} in experiments 
	using Germanium detectors or bolometers. 
	It will be difficult to probe evidence for this rare decay 
	mode in experiments, which have to work - as result of their 
	limited resolution - with energy windows around 
	Q$_{\beta\beta}$ of several hundreds of keV, such as NEMO III, %
	EXO, %
	CAMEO. %

	Another important point is the efficiency 
	of a detector for detection of a $\beta\beta$ signal.
	For example, with 14$\%$ efficiency a potential 
	future 100\,kg $^{82}{Se}$ NEMO experiment would be, because 
	of its low efficiency, equivalent only to a 10\,kg 
	experiment (not talking about the energy resolution).

	In the first proposal for a third generation double 
	beta experiment, the GENIUS proposal 
\cite{KK-Bey97,KK-InJModPh98,KK-H-H-97,KK-Neutr98,KK-00-NOON-NOW-NANP-Bey97-GEN-prop,GEN-prop},
	the idea is to use 'naked' Germanium detectors in a huge tank 
	of liquid nitrogen. It seems to be at present the {\it only} 
	proposal, which can fulfill {\it both} requirements 
	mentioned above - to increase the detector mass 
	and simultaneously reduce the background drastically. 
	GENIUS would - with only 100\,kg of enriched $^{76}{Ge}$ - 
	increase the confidence level of the present pulse shape 
	discriminated 0$\nu\beta\beta$ signal to 4$\sigma$ within 
	one year, and to 7$\sigma$ within three years of measurement 
	(a confirmation on a 4$\sigma$ level by the MAJORANA project 
	would need according to our estimate 
	at least $\sim$230\,years, the CUORE project might 
	need - ignoring for the moment the problem of identification 
	of the signal as a $\beta\beta$ signal - 3700 years).
	With ten tons of enriched $^{76}{Ge}$ GENIUS should be capable 
	to investigate also whether the neutrino mass mechanism 
	or another mechanism (see, e.g. 
\cite{KK60Y})
	is dominating the \znbb~ decay amplitude.


\vspace{-0.3cm}
\section{GENIUS-TF}

	As a first step of GENIUS, a small test facility, GENIUS-TF, 
	is under installation in the Gran Sasso 
	Underground Laboratory 
\cite{GenTF-0012022,NIM02-TF-Elektr} 
	since March 2001.
	With up to 40 kg of natural Ge detectors operated 
	in liquid nitrogen, GENIUS-TF could test the DAMA seasonal 
	modulation signature for dark matter 
\cite{KK-Modul-NIM03}. 
	No other experiment running like, CDMS, IGEX, etc., 
	or projected at present, will have this potential 
\cite{KK-LP01}.
	Up to summer 2001, already six 2.5\,kg Germanium detectors with 
	an extreme low-level threshold of $\sim$500\,eV have been produced.

	The idea of GENIUS-TF is to prove the feasibility 
	of some key constructional features of GENIUS, such as detector 
	holder systems, achievement of very low thresholds 
	of specially designed Ge detectors, long term stability 
	of the new detector concept, reduction of possible noise 
	from bubbling nitrogen, etc.

\begin{figure}[htp]
\centering{
\includegraphics*[scale=0.35]{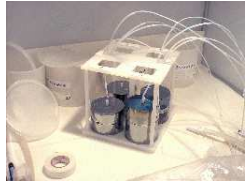}}
\caption[]{\underline{Left:} 
	The first four naked Ge detectors before installation
	into the GENIUS-TF setup. 
	 \underline{Right:}
	Taking out the crystals from the transport dewars 
	and fixing the electrical contacts in the clean room 
	of the GENIUS-TF building - from left to right: 
	Herbert Strecker, Hans Volker Klapdor-Kleingrothaus, Oleg Chkvorez.
\label{fig:Foto-4det}}
\end{figure}

	After installation of the GENIUS-TF setup between halls A 
	and B in Gran Sasso, opposite to the buildings of the \HM 
	double beta decay experiment and of the DAMA experiment, 
	the first four detectors have been installed 
	in liquid nitrogen on May 5, 2003 and have started operation  
\cite{KK-IK-Pusk-GTF,Oleg} 
	(Fig.%
\ref{fig:Foto-4det}).


	This is the first time ever, that this novel technique 
	for extreme background reduction in search for rare decays 
	is tested under realistic background conditions in an underground 
	laboratory.


\vspace{-0.3cm}
\section{Conclusion}
	
	The status of present double beta decay search 
	has been discussed, and 	
	recent evidence for a non-vanishing 
	Majorana neutrino mass obtained by the HEIDELBERG-MOSCOW 
	experiment has been presented. 	
	Additional support for this evidence has been presented  by
	showing consistency of the result - for the signal,  
\mbox{a n d}  
	for the background - with other double beta decay 
	experiments using non-enriched or enriched Germanium detectors. 
	In particular it has been
	shown that the lines seen in the vicinity of the signal (including
	those which at present cannot be attributed) are seen also in
	the other experiments. This is important 
	for the correct treatment of the
	background. Furthermore, the sensitivity of the peak identification
	procedures has been demonstrated by extensive statistical simulations.
	It has been further shown by new extensive simulations of the expected
	background by GEANT4, that the background around 
	$Q_{\beta\beta}$ should be flat, and
	that no known gamma line is expected at the energy 
	of $Q_{\beta\beta}$.
	The 2039\,keV signal is seen  \mbox{o n l y}  in
	the HEIDELBERG-MOSCOW  experiment, which has a {\it by far larger} 
	statistics than all other double beta experiments.

	The importance of this first evidence for violation 
	of lepton number and of the Majorana nature of neutrinos 
	is obvious. It requires beyond Standard Model Physics 
	on one side, and 
	may open a new era in space-time structure 
\cite{AHLUW02}.
	It has been discussed that the Majorana nature of the neutrino 
	tells us that spacetime does realize a construct 
	that is central to construction of supersymmetric theories.

	With the successful start of operation of GENIUS-TF 
	with the first four naked Ge detectors in liquid nitrogen 
	on May 5, 2003 in GRAN SASSO, which is described in 
\cite{KK-Modul-NIM03,KK-IK-Pusk-GTF} 
	a historical step has been achieved of a novel technique and into 
	a new domain of background reduction in underground 
	physics in the search for rare events.

	Future projects to improve 
	the present accuracy of the effective neutrino mass have 
	been briefly discussed. The most sensitive of them and perhaps  
	at the same time most realistic one, is the GENIUS project.
	 GENIUS is the only of the new projects 
	which simultaneously has a huge potential for 
      cold dark matter search, and for real-time detection of 
      low-energy neutrinos (see 
\cite{KK-InJModPh98,KK-Bey97,KK-NOW00,BedKK-01,KK-SprTracts00,KK60Y,KK-IK,KK-LowNu2,KK-NANPino00}).


\vspace{-0.3cm}
        

\end{document}